\documentclass[10pt]{sty/iopart}

\usepackage{sty/iopams}  
\usepackage{xcolor}
\usepackage[title]{appendix} 
\usepackage{cite} 

\usepackage{amsmath}

\usepackage{float}
\usepackage{graphicx}
\usepackage{caption}
\usepackage{subcaption}
\usepackage{appendix}
\usepackage{abstract}
\usepackage{esint}

\usepackage[bookmarks=ture,breaklinks,colorlinks,linkcolor=black,citecolor=red]{hyperref}

\DeclareGraphicsExtensions {.eps ,.mps ,.pdf ,.jpg ,.png}
\graphicspath{{figs/}}

\newcommand{\pp}[2]{\frac{\partial #1}{\partial #2}}

\newcommand{\dd}[2]{\frac{d #1}{d #2}}

\newcommand{\grad}{\nabla}
\renewcommand{\div}{\nabla \cdot}
\newcommand{\curl}{\nabla \times}
\newcommand{\x}{\times}
\newcommand{\cd}{\cdot}

\newcommand{\bbr}[1]{\left( #1 \right)}
\newcommand{\bb}[1]{\left[ #1 \right]}

\newcommand{\vb}[1]{\left.  #1 \right|}
\newcommand{\bd}[1]{\mathbf{#1}}

\usepackage{geometry}[0.8]

\providecommand{\keywords}[1]
{
    \small	
    \textbf{\textit{Keywords---}} #1
}

\begin{document}

    \title[Nonlinear Saturation of Ballooning Modes in Stellarators]{Nonlinear Saturation of Ballooning Modes in Stellarators}

    \author{X. Chu$^{1,2}$, S. C. Cowley$^{1,2}$, N. Ferraro$^{1}$, Y. Zhou$^{3}$, F. I. Parra$^{1,2}$}

    \address{$^1$Princeton Plasma Physics Laboratory, 100 Stellarator Road, Princeton, NJ 08540, USA}
    \address{$^2$Princeton University, Princeton, NJ 08544, USA}
    \address{$^3$School of Physics and Astronomy, Institute of Natural Sciences, and MOE-LSC, Shanghai Jiao Tong University, Shanghai 200240, China}

    \ead{xchu@pppl.gov}
    \address{ORCID IDs:\\
        Xu Chu: https://orcid.org/0000-0002-0382-6428 \\
    }
    \vspace{10pt}
    \begin{indented}
        \item[] July 2023
    \end{indented}
    \begin{abstract}
        Ballooning mode saturation is investigated in realistic stellarator configurations using the flux tube approach of Ham et. al.\cite{ham_nonlinear_2018}\cite{ham_nonlinear_2016}.  
        The method is adapted to account for the lack of exact force balance in stellarator equilibrium solvers that assume existence of nested flux surfaces.
        A variational approach for calculating flux tube energy is developed to overcome this force error problem 
        in stellarator numerical equilibria. Saturated (equilibrium) flux tube states that cross 10-20\% of the plasma minor radius are shown to exist for linearly ballooning unstable profiles.   
        It is shown that several features of the displaced flux tube structure in a full nonlinear MHD simulation of Wendelstein 7X are reproduced by our model. 
        Saturated states are found in a compact stellarator equilibrium close but below the marginal ballooning linear instability, i.e. the unperturbed 
        equilibrium is metastable. 
        This suggests that Edge-Localized-Mode-like explosive MHD behavior may be possible in stellarators. 
    \end{abstract}

    \maketitle
    \ioptwocol

    \section{Introduction}\label{sec:1}

Stellarators usually have better MHD stability properties than tokamaks due
to their small net toroidal current. High $\beta$ operation 
with average $\beta$ from 3\% to 5\% has been demonstrated in various devices such as 
W7AS\cite{hirsch_major_2008}, LHD\cite{sakakibara_mhd_2008} and W7X\cite{klinger_overview_2019}
.
In contrast to tokamaks, where linear MHD stability analysis predicts well the $\beta$ limits,
stellarators can sometimes operate beyond usual linear stability boundary like the Mercier interchange limit\cite{weller_significance_2006}, or the ideal ballooning 
limit\cite{weller_significance_2006}. 
In these cases, magnetic fluctuations are observed after $\beta$ exceeds these linear stability limits, but their amplitude remains small and they merely lead to
an increase in transport without total loss of confinement. This is known as a soft $\beta$ limit, in opposition to hard $\beta$ limits. Hard $\beta$ limits are characterized by events that end in bursty transport with large heat flux to the plasma facing components, usually intolerable in reactor conditions. Exmaples of hard beta limits are disruptions and Edge Localized Modes (ELMs)\cite{leonard_edge-localized-modes_2014} in tokamaks. We need to understand which configurations have soft or hard beta limits in stellarators, motivating the use of nonlinear MHD models. Hard limits can be particularly dangerous for future stellarator power plants where the stored energy is much higher than in stellarators today.

In this paper we focus on the nonlinear behavior of ideal ballooning mode.
Balooning modes are high-mode-number pressure-driven ideal MHD instabilities
that extend along magnetic field lines\cite{connor_high_1979}. They can lead to both soft 
and hard $\beta$ limits. An example of hard $\beta$ limit related to ballooning modes is ELMs in tokamaks.
In stellarators, MHD phenomena resembling ballooning modes has been observed during high $\beta$ operation of W7AS\cite{geiger_equilibrium_2004}.
In addition, a ballooning mode is suspected to be responsible for core density collapse in LHD\cite{ohdachi_observation_2017}.

The linear behavior of ballooning modes has been studied extensively in 
tokamaks \cite{connor_shear_1978}\cite{greene_second_1981} and stellarators\cite{dewar_ballooning_1983}\cite{hegna_stability_1998},
both analytically and numerically \cite{sanchez_ballooning_2000}\cite{miller_hot_1987}\cite{anderson_methods_1990}. 
For the nonlinear behavior of ballooning modes, theory has been developed for the early nonlinear 
stage \cite{cowley_explosive_1997}\cite{hurricane_nonlinear_1997}\cite{henneberg_interacting_2015} and the intermediate nonlinear stage\cite{zhu_ballooning_2008}
when explosive growth of ballooning modes forms isolated plasma `fingers'. A model for the nonlinear saturated stage of ballooning modes
has predicted that certain tokamak configurations are metastable to ballooning modes\cite{ham_nonlinear_2018}\cite{ham_nonlinear_2016}. 

With the advances in extended MHD codes such as M3D-C1, JOREK and NIMROD, the nonlinear behavior of ballooning modes
can be studied numerically. The recent extension of M3D-C1 to stellarator geometries\cite{zhou_approach_2021} has enabled the study of 
W7X high beta configurations near the ideal ballooning limit where benign saturation is observed\cite{zhou_benign_2024}.

In this work, we study the nonlinear saturation of ballooning modes in stellarators by implementing 
the flux tube model described in \cite{ham_nonlinear_2018} in realistic stellarator numerical equilibria obtained
with the DESC equilibrium solver\cite{panici_desc_2023}. 
We show that saturation levels can be calculated and that metastable
stellarator equilibria exist. These metastable equilibria might explain
ELM-like behavior that has been observed in stellarators\cite{hirsch_major_2008}.

The paper is organized as follows. In Section \ref{sec:2}, the flux tube model for ballooning modes 
developed in \cite{ham_nonlinear_2018} is reviewed and 
the numerical implementation in real stellarator geometries is presented.
In Section \ref{sec:3}, saturated ballooning modes are qualitatively described. Convergence analysis and benchmarks are 
presented. The model is then compared with the flux tube structure observed in a recent M3DC1 simulation for W7X\cite{zhou_benign_2024} 
in the nonlinear saturation stage, The assumptions used in the model are tested and 
the flux tube shape in the simulation is compared with the model, showing good agreement.
In Section \ref{sec:4}, energy release is calculated. 
In Section \ref{sec:5}, a case study of a compact QA(Quasi-axisymmetric) equilibrium\cite{nelson_design_2002} near ideal ballooning marginal stability 
is presented and the existence of metastable equilibria is demonstrated.
In Section \ref{sec:7}, we discuss the applicability of the model and how connections might be drawn to soft/hard $\beta$ limits in stellarators.
    \section{Flux Tube Model and Numerical Implementation}\label{sec:2}

In this section we revisit the model in \cite{ham_nonlinear_2018}, and apply it to general 3D configurations.

\begin{figure}[htbp]
    \centering
    \includegraphics[width=.45\textwidth]{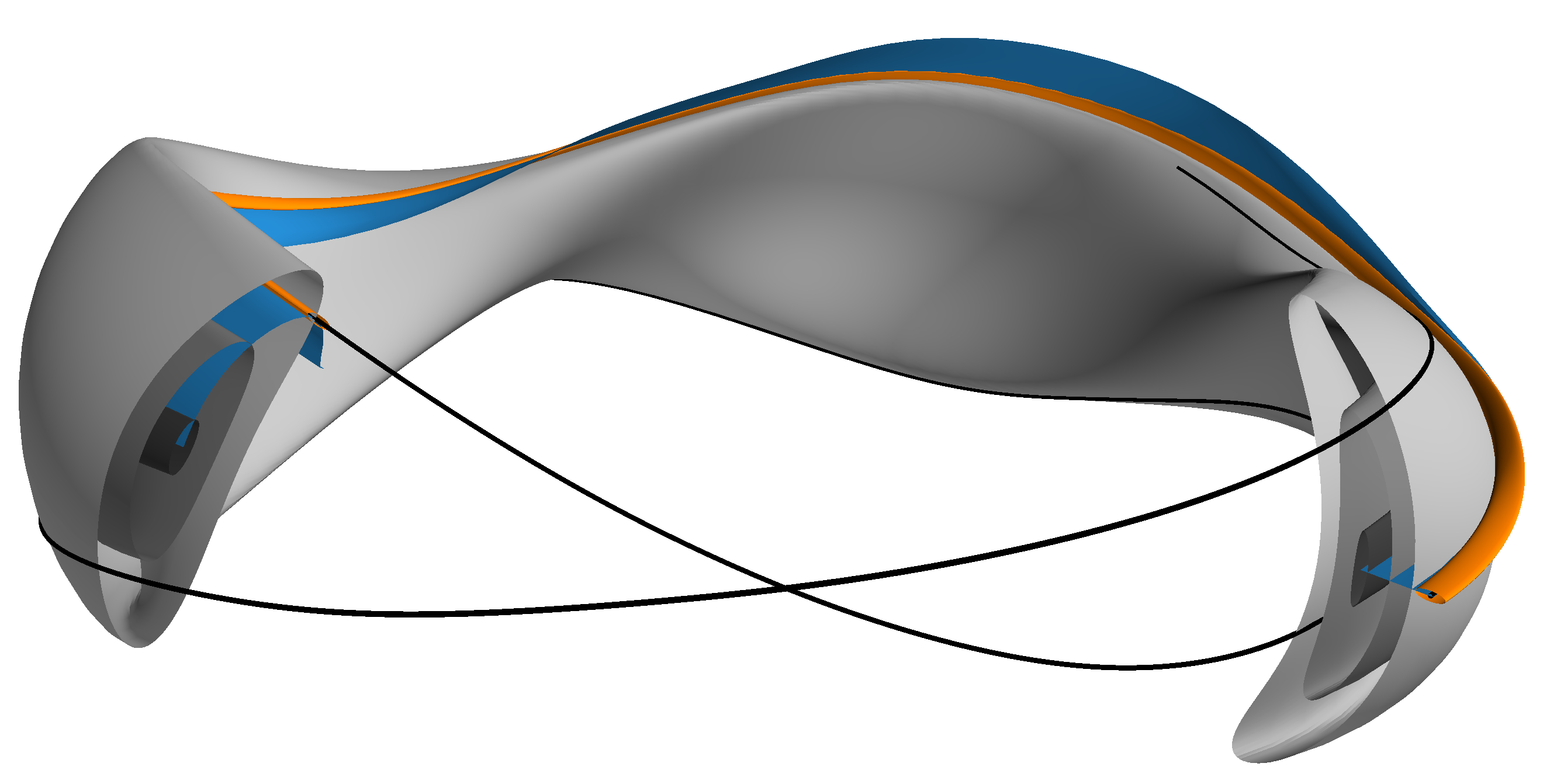}
    \caption{An illustration of the geometry of the flux tube model using a compact QA\cite{nelson_design_2002} equilibrium.
        Flux surfaces are plotted in grey. The surface of constant $\alpha$, which is tangential to 
        the equilibrium field at all points, is depicted in blue. 
        The perturbed flux tube (orange color) moves on the $\alpha$ surface with its
        elliptical cross-section elongated along the $\alpha$ surface. 
        A field line inside the flux tube is plotted in black -- 
        note that it is bent with respect to the equilibrium field.  
        We assume that the flux tube does not intersect itself after going around the device. }
    \label{fig:Model}
\end{figure}

We consider a perturbed state of the equilibrium, 
with a field line aligned flux tube displaced from its original position, as shown in Fig.\ref{fig:Model}. 
The flux tube is thin and enlongated in the direction of its movement,\cite{ham_nonlinear_2018} 
so that the change of the background magnetic field remains unimportant as the flux tube slides along a 
surface that is everywhere tangential to the background magnetic field.  
We call such a surface an \textbf{$\alpha$ surface} - see Eq. \eqref{eq:alphasurfcedef} and the discussion below for more details.  
By moving along an $\alpha$ surface, 
the perturbed flux tube does not cross any background magnetic field lines, i.e. the topology of the field lines does not change.  
The background field lines simply part to allow the perturbed flux tube to pass.  

The magnetic field and pressure inside the flux tube are denoted $\mathbf{B}_{in}$ and $p_{in}$,
respectively. Just outside the tube, the magnetic field and the pressure are 
$\mathbf{B}_{out}$ and $p_{out}$.  
We assume the flux tube is sufficiently narrow that it does not perturb the magnetic field nor 
the pressure outside the tube. Thus,
\begin{align}
    \mathbf{B}_{out}=\mathbf{B}_{0} = \grad \psi \x \grad \alpha_{0}, \;\; p_{out} = p_0{(\psi)},
\end{align}
where $\mathbf{B}_{0}$ is the original equilibrium field, $p_0(\psi)$ is the equilibrium pressure profile
and $\psi$ is the toroidal magnetic flux divided by $2\pi$. The quantity $\alpha_{0} = \theta - \iota \zeta$ is 
the nonperiodic Clebsch coordinate,  
where $\zeta$ is an arbitrary toroidal angle, $\theta$ is a corresponding straight field line poloidal 
angle, and $\iota=\iota(\psi)$ is the rotational transform. 
Since DESC\cite{panici_desc_2023} is used for equilibrium calculation, 
$\zeta$ from here on should be understood as the toroidal angle in cylindrical coordinates. 
However, the theory can be applied to any general toroidal angle without any modifications.

Since $\mathbf{B}_{0}\cdot\grad\alpha_{0} = \mathbf{B}_{0}\cdot\grad\psi =0$, 
a general surface that is everywhere tangent to the background equilibrium magnetic field can be written as
\begin{align}\label{eq:alphasurfcedef}
    \alpha = \alpha_{0} - f(\psi) =0,
\end{align}
where the flux function $f(\psi)$ is undetermined to the order that we perform these calculation.
We refer to the surface determined by Eq. \eqref{eq:alphasurfcedef} as \textbf{$\alpha$ surface} from here on. 
Note that $\mathbf{B}_{0} = \grad \psi \x \grad \alpha$ for all $f(\psi)$.
Choosing an $\alpha$ surface is equivalent to choosing $f(\psi)$.  
The alpha surface intersects itself on rational surfaces when traced around the machine one or more times.  
However, we assume that the perturbed flux tube does not intersect itself where the perturbation is significant.  
In such cases, we can locate points on the alpha surface by using the coordinates of $\psi$ and $\zeta$.

The tangent space of the $\alpha$ surface is spanned by $\mathbf{b}_0$ and $\mathbf{e}_\perp$, 
where $\mathbf{b}_0=\mathbf{B_0}/|\bd{B_0}|$ is a unit vector in the direction of the background magnetic field, 
and $\mathbf{e}_\perp$ is a vector perpendicular to $\mathbf{b}_0$ and $\grad \alpha$,
\begin{align}
    \mathbf{e}_\perp = \grad \alpha \x \mathbf{b}_0.
\end{align}
All equilibrium quantities ({\it e.g.} $ \mathbf{e}_\perp, \mathbf{b}_0, \grad\alpha, B_0$) are known on the alpha surface as functions of $\psi$ and $\zeta$.  
To simplify our notation, the $\alpha$ dependence is not made explicit for the rest of this paper.  

The perturbed flux tube has a width $\delta_\alpha$ in the $\grad\alpha$ direction, 
and a width of $\delta_\perp$ along the $\alpha$ surface in the $ \mathbf{e}_\perp$ direction.  
We assume $\delta_\alpha \ll \delta_\perp\ll L$ where $L$ is the typical equilibrium scale -- for a more detailed 
discussion of the limit of validity of this flux tube model, see  \cite{ham_nonlinear_2018}.  
Each field line inside the flux tube stays on the alpha surface (to leading order in $\delta_\alpha/L \ll 1$) 
and is represented by its radial ($\psi$) displacement (with respect to its initial location) $\eta$ as a function of 
the position along the line, given by the toroidal angle $\zeta$,
\begin{align}
    \psi = \psi_0 + \eta(\zeta, \psi_0)
\end{align}
Here, $\psi_0$ is the flux surface label where the field line originally locates.  
We assume that the field lines inside the tube are bent with a radius of curvature of the order of the equilibrium length scale; 
thus $\partial\eta / \partial\zeta \sim{\cal{O}}(\psi_{edge})$, where $\psi_{edge}$ is the 
value of normalized toroidal flux $\psi$ at the edge of the stellarator.
Since the magnetic field inside the flux tube is tangent to the $\alpha$ surface, it can be represented as 
\begin{align}\label{eq:Bin_rep}
    \mathbf{B}_{in}(\zeta, \psi_0) = B_\parallel(\zeta, \psi_0) \mathbf{B}_0(\psi, \zeta) + B_\perp(\zeta, \psi_0) \mathbf{e}_\perp(\psi, \zeta).   
\end{align}
Note that $\mathbf{B}_0$ and $\mathbf{e}_\perp $ are evaluated at the perturbed location $\psi=\psi_0 + \eta(\zeta)$ on the $\alpha$ surface.
We can relate $B_\perp$ to $\eta$ and its derivative,
\begin{align}\label{eq:ode1-fl}
    \vb{\dd{\eta}{\zeta}}_{\psi_0} = \frac{\mathbf{B}_{in}\cd\grad\psi}{\mathbf{B}_{in}\cd\grad\zeta} 
    = \frac{B_\perp \mathbf{e}_\perp  \cd \grad \psi }{ B_\parallel \mathbf{B}_0\cd \grad \zeta + B_\perp \mathbf{e}_\perp  \cd \grad \zeta }.
\end{align}

The equilibrium state of the flux tube is obtained from MHD force balance\cite{ham_nonlinear_2018}. From parallel force balance ($\mathbf{B}_{in}\cd\grad p_{in} =0$), 
we find that hydrodynamic pressure is constant to lowest order along the perturbed field line. 
Since the perturbed flux tube connects to the initial surface $\psi_0$ at $\zeta \to \infty$, $p_{in}=p_0(\psi_0)$.  
The smallest spatial scale of the flux tube, $\delta_\alpha$, is in the $\alpha$ direction. 
In the perpendicular force balance in that direction, the magnetic field line tension can be neglected to leading order compared to
the total pressure  ($p + B^2/(2\mu_0)$) gradient, where $\mu_0$ is the vacuum permittivity.
As a result, $B_{in}=|\mathbf{B}_{in}|$ is known as a function of spacial coordinates alone,
\begin{align}
    B_{in}^2 &= B_{out}^2 + 2\mu_0 (p_{out}-p_{in})\nonumber \\ &= B_{0}^2(\psi, \zeta) + 2\mu_0 (p_{0}(\psi)-p_0(\psi_0)),
\label{eq:Bin}\end{align}
where $B_{out}=B_{0}(\psi, \zeta)$ and $p_{out}=p_0(\psi)$ are evaluated at the perturbed location of the flux tube. We can therefore relate $B_\parallel$ to $B_\perp$ as
\begin{equation}\label{eq:totalpresBalalnce}
    B_\parallel = \sqrt{1 + 2\mu_0\Delta p/B_{0}^2 - |\mathbf{e}_\perp |^2B_\perp^2/B_{0}^2 },
\end{equation}
where $\Delta p = p_{0}(\psi)-p_0(\psi_0)$.  

Using the results above and force balance in the $\mathbf{e_\perp}$ direction, we have an equation for $\eta$.
Using Eq. (\ref{eq:Bin}) and the background magnetic field force 
balance $ \mathbf{B}_0 \cdot \grad \mathbf{B}_0 - \grad (p_0 + B_0^2/2\mu_0) = 0$, and 
imposing that the force along $\mathbf{e_\perp}$ on the perturbed flux tube vanishes give the generalised {\it Archimedes principle} equation 
\begin{align}
    \mu_0 F_\perp = &(\mathbf{B}_{in}\cd\grad \mathbf{B}_{in} -  \mathbf{B}_0\cd\grad\mathbf{B}_0)\cd\mathbf{e}_\perp =0.
\end{align}
Using the representation of the $\mathbf{B}_{in}$ in Eq. (\eqref{eq:Bin_rep}), this equation can be expanded as  
\begin{align}\label{eq:intermed}
    &\mathbf{B}_{in}\cd \grad (|\mathbf{e}_\perp|^2 B_\perp) + (B_\parallel^2-1) B_0^2 \boldsymbol{\kappa}\cd \mathbf{e}_\perp \nonumber\\ 
    &+ B_\perp B_\parallel \bbr{\mathbf{e_\perp} \cd \grad \mathbf{B_0}} \cd \mathbf{e_\perp} - B_\perp\bbr{\mathbf{B}_{in} \cd \grad \mathbf{e}_\perp} \cd \mathbf{e}_\perp= 0.
\end{align}
Note that we have the following equalities
\begin{align}\label{eq:notice}
    \bbr{\mathbf{e_\perp} \cd \grad \mathbf{b_0}} \cd \mathbf{e_\perp} &= - \mathbf{b}_0 \cd \grad |\mathbf{e}_\perp|^2/2 + 
    \bb{\mathbf{e_\perp} \x \bbr{\curl \mathbf{e}_\perp} } \cd \mathbf{b}_0\nonumber \\ 
    &=  \mathbf{b}_0 \cd \grad |\mathbf{e}_\perp|^2/2 - |\mathbf{e}_\perp|^2 \mathbf{b}_0 \cd \grad B_0/B_0.
\end{align}
Here, the definition $\mathbf{e_\perp} = \grad \alpha \x \mathbf{b}_0$, $ \mathbf{b}_0 \cd \grad \alpha= 0$, $|\bf{e_\perp}|^2=|\grad \alpha|^2$ 
and $\div \bf{B_0} = 0$ are utilized to simplified the second term on the RHS of the first equality.
Substituting Eq. \eqref{eq:notice} into Eq. \eqref{eq:intermed}, we derive the nonlinear ballooning equation
\begin{align}\label{eq:forcebalance}
    \mathbf{B}_{in}\cd \grad &\bbr{\frac{|\mathbf{e}_\perp |^2}{B_{0}}B_\perp} + (B_\parallel^2-1) B_{0} \boldsymbol{\kappa}\cd \mathbf{e}_\perp  
    \nonumber\\ &- B_\perp^2 B_{0} \mathbf{e}_\perp  \cd \grad \bbr{\frac{|\mathbf{e}_\perp |^2}{2B_{0}^2}} = 0.
\end{align}
where $\boldsymbol{\kappa} = \mathbf{b}_0\cd\grad\mathbf{b}_0$ is the background magnetic field curvature.
For convenience, we introduce another variable $Y$,
\begin{align}
    Y(\zeta, \psi_0) = \frac{|\mathbf{e}_\perp |^2}{B_{0}}B_\perp.
\label{eq:Ydef}\end{align}
We note that $\mathbf{B}_{in}\cd \grad Y = ({\mathbf{B}_{in}\cd \grad \zeta})\vb{\dd{Y}{\zeta}}_{\psi_0}$. Thus,
\begin{align}
    \vb{\dd{Y}{\zeta}}_{\psi_0} &= \frac{B_0 }{ B_\parallel \mathbf{B}_0\cd \grad \zeta + B_\perp \mathbf{e}_\perp  \cd \grad \zeta } 
    \bigg[ B_\perp^2  \mathbf{e}_\perp  \cd \grad \bbr{\frac{|\mathbf{e}_\perp|^2}{2B_{0}^2}}
    \nonumber\\
    & - (B_\parallel^2-1) \boldsymbol{\kappa}\cd \mathbf{e}_\perp  \bigg].
\label{eq:ode2}\end{align}

Eqs.~(\ref{eq:ode1-fl}) and (\ref{eq:ode2}) along with the algebraic relations Eqs.~(\ref{eq:totalpresBalalnce}) and (\ref{eq:Ydef}) constitute two coupled first-order nonlinear ordinary 
differential equations for $\eta(\zeta, \psi_0)$ and $Y(\zeta, \psi_0)$.  The right hand side of Eqs.~(\ref{eq:ode1-fl}) and (\ref{eq:ode2}) can be evaluated algebraically given the background MHD equilibrium,
$\eta$, $Y$ and the $\alpha$ surface.  Note that $\psi_0$ is a parameter in these equations and thus each field line in the flux tube can be solved independently. 
In this paper we assume that the tube is small enough that one field line is representative of the motion of the whole flux tube and we do not attempt to solve for the structure inside the tube.

We seek solutions of Eqs.~(\ref{eq:ode1-fl}) and (\ref{eq:ode2}) with $\eta \rightarrow 0$ and $Y\rightarrow 0$ as $\zeta\rightarrow \pm \infty$.  
The solution for a chosen $\psi_0$ can in principle be found via a shooting method.
We chose a simulation domain extended along the 
field line, given by $\zeta \in [\zeta_0, \zeta_1]$, where $\zeta_0$ and $\zeta_1$ are large numbers, usually encompassing six or more toroidal turns.
At $\zeta_0$, we impose zero displacement and we provide a guess $Y_0$ for the value of $Y$.
Eqs.~(\ref{eq:ode1-fl}) and (\ref{eq:ode2}) are then integrated numerically to obtain $Y(\zeta, \psi_0)$ and $\eta(\zeta, \psi_0)$
for $\zeta \in [\zeta_0, \zeta_1]$.
In general, for any given guess $Y_0$, $\eta$ does not go to zero at $\zeta_1$, but
for some equilibria and particular nonzero discrete values of the amplitude $Y_0$, the numerical solution satisfies the boundary condition $\eta(\zeta_1) = 0$.
If such nonzero values of $Y_0$ exist then these are valid saturated flux tube states.
We use the brentq root finding algorithm\cite{brent_algorithms_1972-1} to obtain these nonzero values of $Y_0$.
First, a set of solutions of Eq. \eqref{eq:forcebalance} is generated using different values of $Y_0$. We use this scan to determine the intervals 
in $Y_0$ that contain zero displacement solutions at the right boundary $\zeta_1$. The interval information is then passed to the brentq algorithm 
to find the accurate $Y_0$ value that satisfies the boundary condition.

We have obtained the solution for $\eta$ for large values of $\zeta$ (\ref{ap:decay}). This analytical solution is not 
used to prescribe the boundary conditions for the shooting method because the derivation assumes that the solution extends sufficiently far along the magnetic 
field line that it samples a significant portion of the flux surface $\psi_0$. This would require extremely long simulation domains.


In \cite{cowley_explosive_2015}, an energy functional for saturated flux tubes was derived.
It is the total magnetic energy in a flux tube per unit magnetic flux,
\begin{align}\label{eq:energy}
    \mathcal{E} = \lim_{x\rightarrow\infty}\left(\int_{-x}^x \mathbf{B}_{in} \cd d\mathbf{r} - \int_{-x}^x \mathbf{B}_{0} \cd d\mathbf{r}_0\right),
\end{align}
where the first functional integral is taken along a perturbed field line (${\bf B}_{in}$) on the $\alpha$ surface, with the field strength ${B}_{in}$ given by Eq.~(\ref{eq:Bin}), and  
the second integral in Eq.~(\ref{eq:energy}) is along the unperturbed field line.
Flux tube equilibrium solutions to Eqs.~(\ref{eq:ode1-fl}) and (\ref{eq:ode2}) are stationary field line paths of the functional ({\it i.e.} paths for which the 
functional derivative vanishes, $\delta {\cal E}/\delta {\bf r}=0$).  Stable flux tube equilibria are (local or global) minima of $\cal E$.

In \ref{ap:energy}, a variational derivation of Eq.~(\ref{eq:forcebalance}) from the energy functional (\ref{eq:energy}) is given. 
It is worth noting that this derivation requires the assumption of equilibrium force balance. 
It turns out that in a numerical equilibrium where only approximate force balance is achieved, the compatibility of this energy functional with the flux tube saturation equation is not guaranteed. 
We will discuss this more in the following sections.

The set of ODEs in Eqs.~(\ref{eq:ode1-fl}) and (\ref{eq:ode2}) is implemented using MHD equilibrium solutions of the DESC equilibrium solver\cite{panici_desc_2023}. 
First, an $\alpha$ surface is selected by choosing $f$ in $\alpha = \theta - \iota \zeta - f(\psi) = 0$. 
Then, a 2D field line aligned grid is generated on the chosen $\alpha$ surface. This grid is uniform in $\rho-\rho_0$ and $\zeta$. Here the flux surface label, $\rho=\rho(\psi)$, 
is the square root of the normalized toroidal flux, and $\rho_0 = \rho(\psi_0)$ is the flux surface label of the unperturbed flux tube. 
Equilibrium-related quantities needed for evaluating the right-hand side of the ODEs are calculated on the grid and interpolated. 
Saturated flux tubes are calculated using the shooting method described above. 
The detailed implementation based on DESC is outlined in \ref{ap:nu}.

For the boundary condition to work, the perturbed flux tube must asymptotically approach the unperturbed flux tube for large $\zeta$.
This requires $\eta \sim \zeta^{\nu}$ with $\nu < -1$ 
as $\zeta \to \infty$ because the displacement in poloidal angle is
$\Delta\theta =  \iota(\psi_0+ \eta(\zeta))\zeta+ f(\psi_0 + \eta(\zeta) ) \approx \zeta \iota' \eta(\zeta) \sim \zeta^{\nu+1}$ due to finite magnetic shear $\iota'$. 
A displacement $\eta \sim \zeta^\nu$ with $\nu \ge - 1$ would cause a poloidal displacement $\Delta \theta$ with respect to the original 
field line that does not decrease as $\zeta$ increases.
The multi-scale asymptotic expansion performed in \ref{ap:decay} predicts the value of $\nu$.
The exponent $\nu$ is the same as the one obtained from the Mercier analysis of the linear ballooning equation as long as $\nu<-1$. Therefore, the calculation is limited to $\nu_{Mercier}<-1$.
For $\nu\ge-1$, the analysis in \ref{ap:decay} is no longer valid because the displaced flux tube does not tend towards the original flux tube for large $\zeta$.
We have found that, for $\nu \ge -1$, numerical solutions to the nonlinear ballooning ODE with zero displacement at the simulation boundary can still be found. 
However, they intersect the original field line at a finite slope and tend to balloon further radially outwards if the numerical domain is extended, that is, for cases with $\nu\ge-1$, the  solutions for the displaced flux tube do not converge with simulation domain size.

    \section{Flux Tube Saturated States and Comparison with M3D-C1}\label{sec:3}

In this section, we first show typical numerical solutions to the nonlinear ballooning equation
\ref{eq:forcebalance} in general stellarator geometry. We also benchmark the code with linear ballooning 
stability calculations. Then, we compare the model with nonlinear MHD simulations. We confirm key assumptions of our model by examining a nonlinear M3D-C1 simulation of a W7-X configuration that has benign saturation of ballooning modes\cite{zhou_benign_2024}, and we compare the flux tube shape provided by our model to the flux tube states extracted from that simulation.

\subsection{Calculation of Flux Tube Saturated States}

\begin{figure}[htbp]
    \centering
    \includegraphics[width=.45\textwidth]{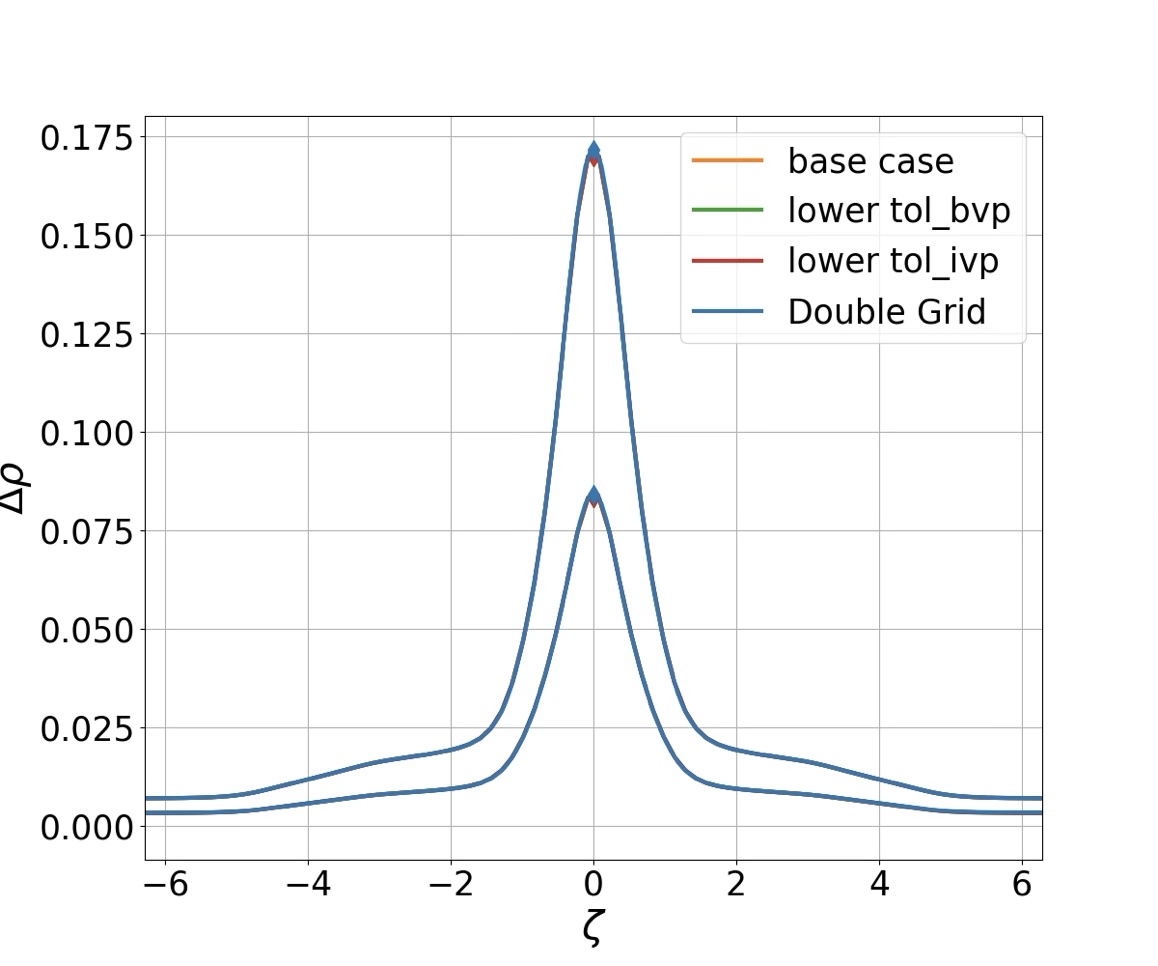}
    \caption{Typical shape of the saturated flux tube states, plotted as radial displacement in normalized minor radius $\Delta \rho$ v.s.
    cylindrical toroidal angles.
    This case is calculated on a compact QA\cite{nelson_design_2002} equilibrium on a stellarator symmetric $\alpha$ surface in the domain $ \zeta \in [-12 \pi, 12 \pi]$. Two saturated
    states are presented, each corresponding to a stationary point of the energy curve.
    The procedure for generating such energy curve will be presented in section \ref{sec:4}.
    Four solutions of different resolutions are plotted for each saturated state, showing good convergence.
    }
    \label{fig:flux-tube-exsample}
\end{figure}

In Fig. \ref{fig:flux-tube-exsample}, examples of the shape of saturated flux tube states are shown. 
In this figure, we plot the change in normalized minor radius
$\Delta \rho = \rho(\zeta) - \rho_0$ as a function of $\zeta$.
There is a discrete number of saturated flux tubes on a given $\alpha$ surface, each corresponding to 
a stationary point in the energy curve. In this particular case, there are two different nonlinear solutions (other than the unperturbed state), 
and the flux tube with the largest displacement is stable, whereas the other flux tube is unstable. 
The stability is determined by the energy dependence on the displacement. The stable flux tube is at a local energy minimum 
and the unstable at a local maximum.
We do not show the energy map here. It will be discussed in detail in section \ref{sec:4}.
The flux tube deforms radially outwards in the region with bad curvature as in the linear ballooning mode calculation. In fact, the shape of the nonlinear saturated flux tube 
qualitatively resemble the linear ballooning eigenmode shape.

The convergence of the flux tube shape has been checked, with respect to 
solution domain size,
interpolation resolution, ODE solving tolerance and root finding tolerance for the shooting method. As shown in the Fig. \ref{fig:flux-tube-exsample}, 
curves obtained with different resolutions overlap. 
For the base case, 500 $\zeta$ grid points for $\zeta\in [-12\pi,12\pi]$ and 41 $\Delta \rho$ grid points for $\Delta \rho \in [-0.1,0.2]$ are used. A relative tolerance of $1\x 10^{-6}$ and an absolute tolerance of $1\x10^{-7}$
are used for the RK45 ODE solver. An absolute tolerance of $2\x 10^{-12}$ and relative tolerance of $1\x10^{-14}$ are used for the root finding algorithm. 
For the "lower tol\_bvp" case, the absolute tolerance of the root finder is improved to $2\x 10^{-14}$. For the "lower tol\_ivp" case, the relative tolerance and the absolute tolerance of the 
RK45 ODE solver are improved to $1\x 10^{-10}$ and $1\x10^{-11}$, respectively. For the double grid case, the grid points in the $\zeta$ direction are increased to 1000.

\begin{figure}[htbp]
    \centering
    \includegraphics[width=.45\textwidth]{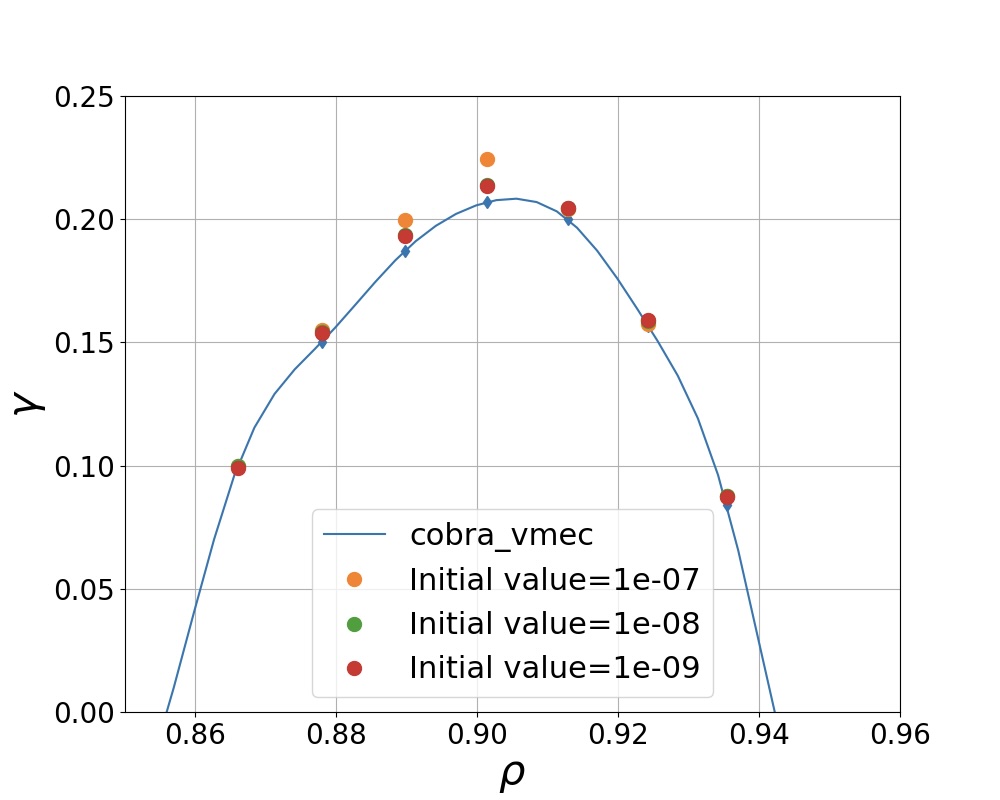}
    \caption{
        Benchmark against COBRAVMEC. Here, $\rho$ is normalized minor radius and $\gamma$ is linear ballooning growth rate normalized to the Alfvén time. 
        The growth rates calculated from our nonlinear model with small initial condition are shown with solid dots.
        The initial values $Y$ have the unit of meters$^{-1}$
    }
    \label{fig:enter-label}
\end{figure}

The nonlinear solver is benchmarked against linear ballooning results from COBRAVMEC. This benchmark is performed by adding the linear inertia term 
$m_i n_i \Delta \rho \omega^2 |\mathbf{e}_\perp|^2/B_0^2 $ on the RHS of Eq. \eqref{eq:forcebalance}.
Then Eq.\ref{eq:forcebalance} is solved with a very small slope starting at one end of the grid, and $\omega^2$ is adjusted
in the shooting method solver to have the solution cross zero at the other side of the domain. The calculated growth rate is 
compared with COBRAVMEC in Fig. \eqref{fig:enter-label}, showing good agreement as the initial slope is decreased.

\subsection{Comparison with Nonlinear MHD Simulation}
To verify the assumptions used in the derivation of this flux tube model and check the accuracy of its prediction,
we proceed to compare our model to a recent ideal MHD M3D-C1 simulation of W7X where benign saturation of ballooning
modes is observed\cite{zhou_benign_2024}. 
We show that the flux tube structures studied in this paper are indeed present in the nonlinear saturation 
stage of the simulation. We check the assumptions made in the theory against the flux tube structure in the simulation, 
and we find that the assumptions are justified. As a result,
the flux tube shape that we aclculate are also in good agreement with the flux tube structure
extracted from the simulation.

\begin{figure}[H]
    \centering
    \includegraphics[width=.5\textwidth]{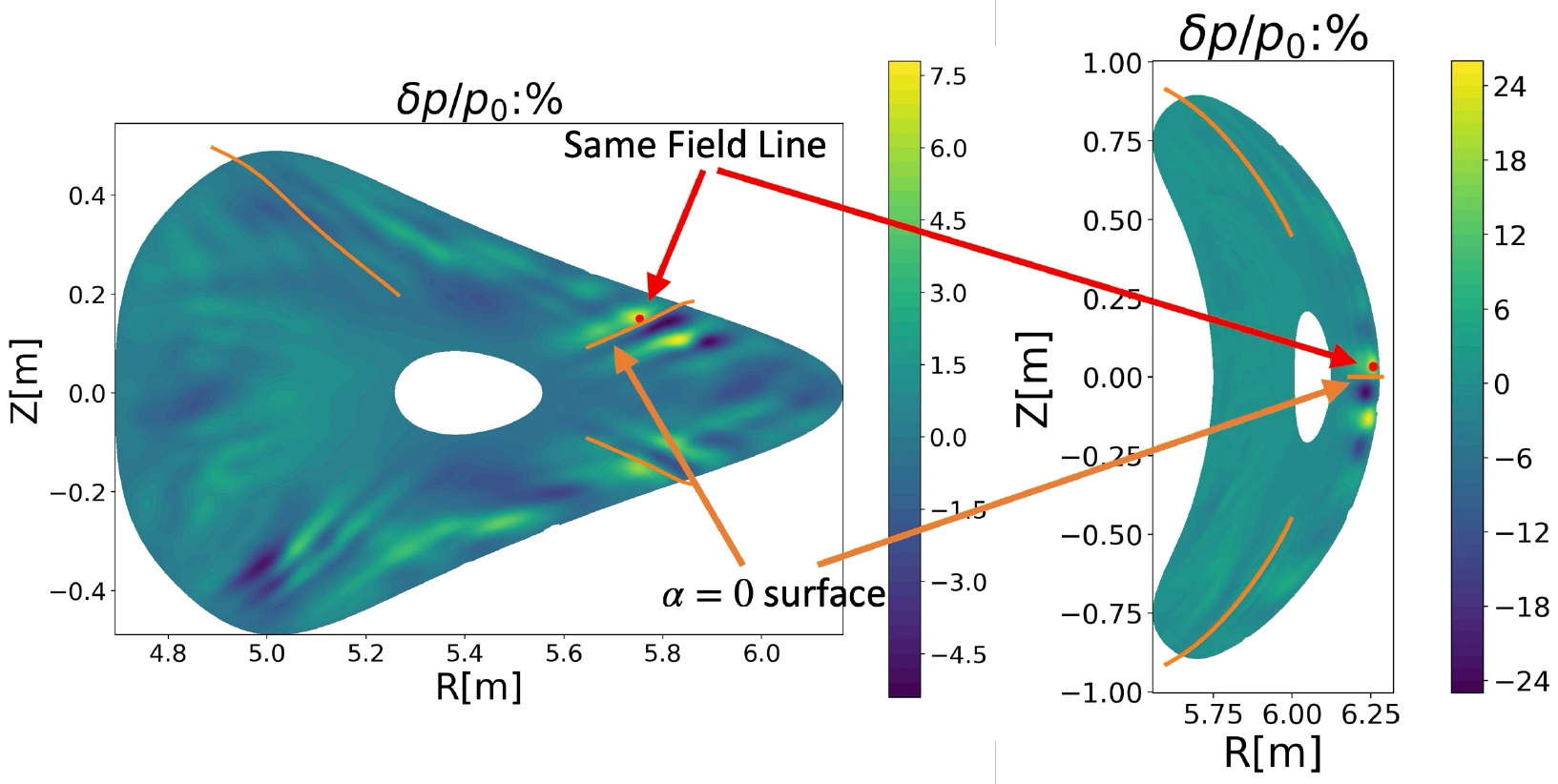}
    \caption{Pressure perturbation on the $\zeta=\pi/5$ (left subfigure) and $\zeta=0$ (right subfigure)
        cross sections at $t=2500 \tau_A$ of the M3D-C1 simulation reported in \cite{zhou_benign_2024}. 
        Red solid circles are the intersection of the same field line with the cross sections.
        The three orange lines are the $\alpha$ surfaces 
        $\theta - \iota \zeta = -2\pi/5, 0, 2\pi/5$.
    }
    \label{fig:W7Xdeltap}
\end{figure}

We examine the time slice $2500\tau_A$ of the two cross sections at $\zeta=0,\pi/5$ 
of the EIM 5.4\% $\beta$ case 
in \cite{zhou_benign_2024}, where ballooning modes have saturated. 
Here $\tau_A=0.46\mu s$ is the Alfvén time, for a reference length of $1m$, a reference magnetic field of $1T$ and a reference density (hydrogen plasmas) of $10^{20} m^{-3}$.
The perturbed pressure $\delta p$ is plotted in Fig. \ref{fig:W7Xdeltap}. 
Here $\delta p = p - p_{0}$ with $p_0$ obtained from the initial VMEC equilibrium and 
$p$ obtained from M3D-C1 simulation of $t=2500\tau_A$.
There are localized peaks of pressure perturbation near the edge of the simulation which 
resemble the saturated flux tube structures that we are studying.
We trace the magnetic field starting at the local maximum of pressure perturbation at 
$\zeta=0$, indicated by the red dot in the right sub-figure. The same magnetic field line
goes through the region of maximum pressure perturbation at $\zeta=\pi/5$, as show in the 
left sub-figure. This confirms that the localized peaks of pressure perturbation are a single flux tube extended toroidally.
At $\zeta=\pi/5$, the localized structure of the pressure perturbation appears to be elongated
radially. This is consistent with our assumption of a flux tube elongated along an $\alpha$ surface. 
The intersections of $\theta - \iota \zeta=0,-2\pi/5,2\pi/5$ surfaces
with the two toroidal cross sections are indicated by the orange curves, and they indeed 
follow the direction of the elongated structure in the pressure perturbation. 

\begin{figure}[H]
    \centering
    \begin{subfigure}[t]{.25\textwidth}
        \centering
        \includegraphics[width=.9\textwidth]{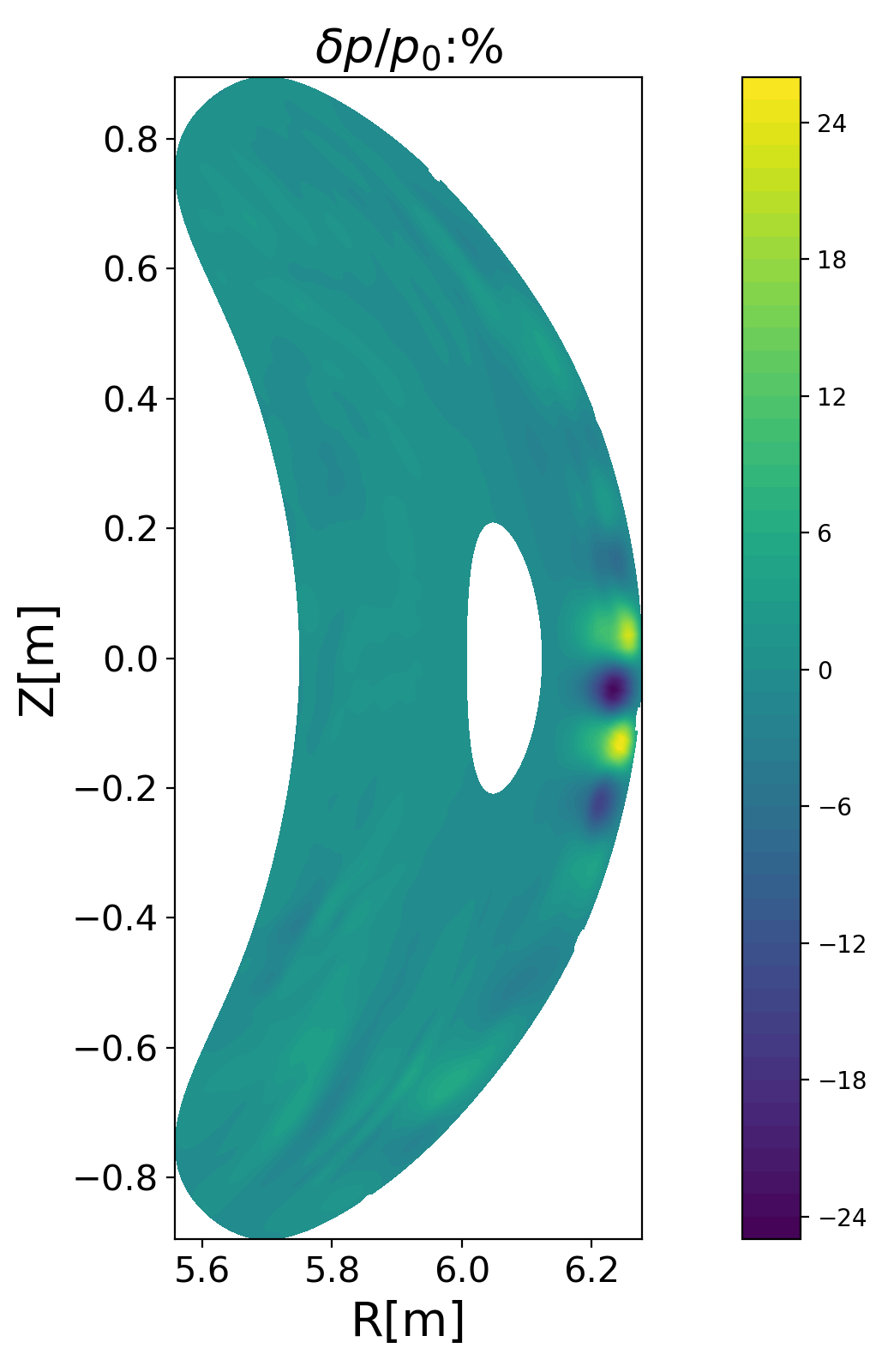}
        \caption{Pressure Perturbation}
    \end{subfigure}%
    ~
    \begin{subfigure}[t]{.25\textwidth}
        \centering
        \includegraphics[width=.9\textwidth]{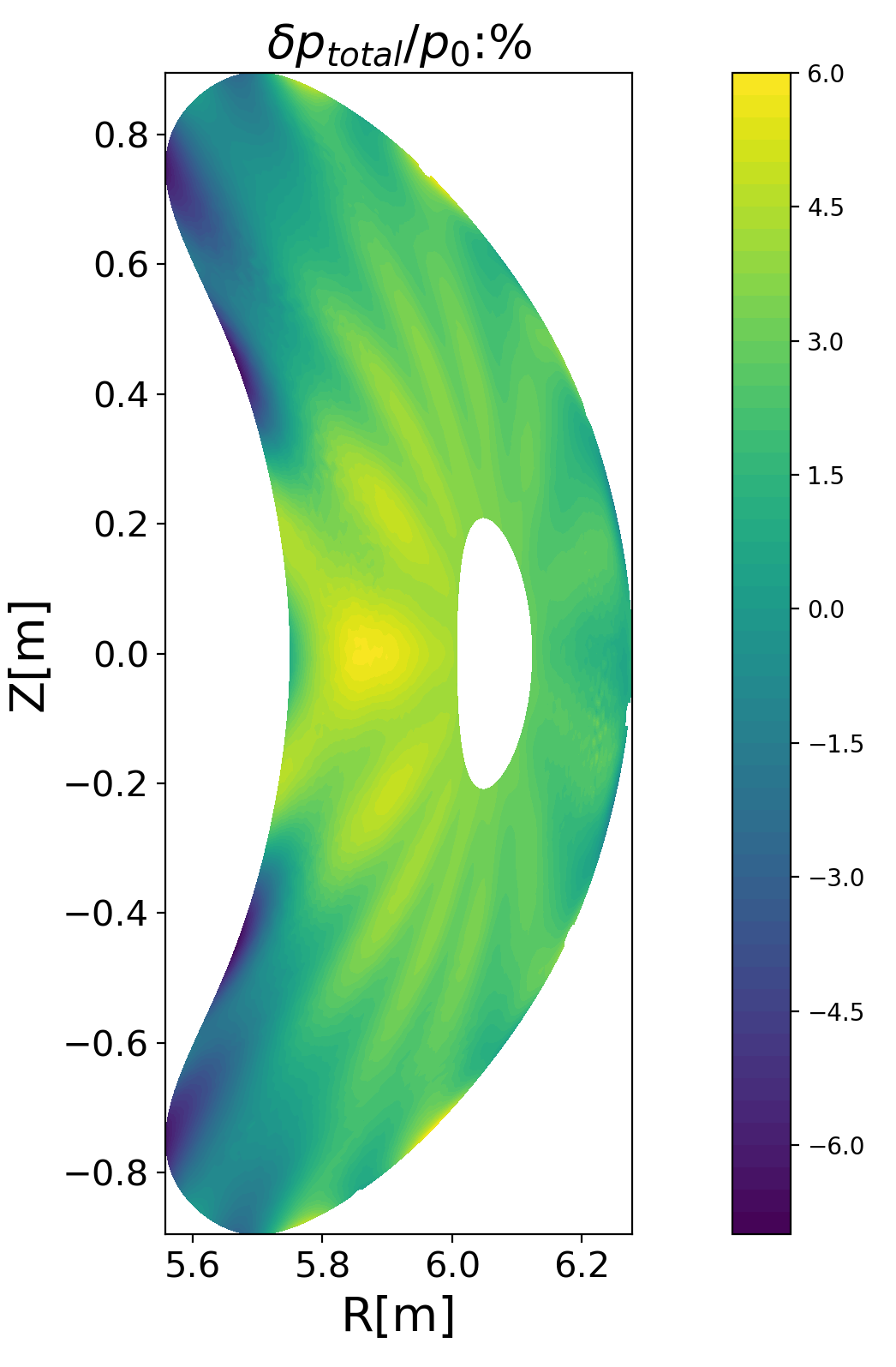}
        \caption{Total Pressure Perturbation}
    \end{subfigure}
    ~
    \caption{Comparison of relative pressure perturbation  $\delta p / p_0$ and relative total pressure perturbation 
    $\delta p_{total} / p_0$, at $\zeta=0, \,t=2500\tau_{A}$. Notice that the small scale structures in $\delta p / p_0$ are not present in $\delta p_{total} / p_0$,
    justifying the assumption of total pressure continuity in our model.}
    \label{fig:W7X_totalp}
\end{figure}

\begin{figure}[htbp]
    \centering
    \includegraphics[width=.45\textwidth]{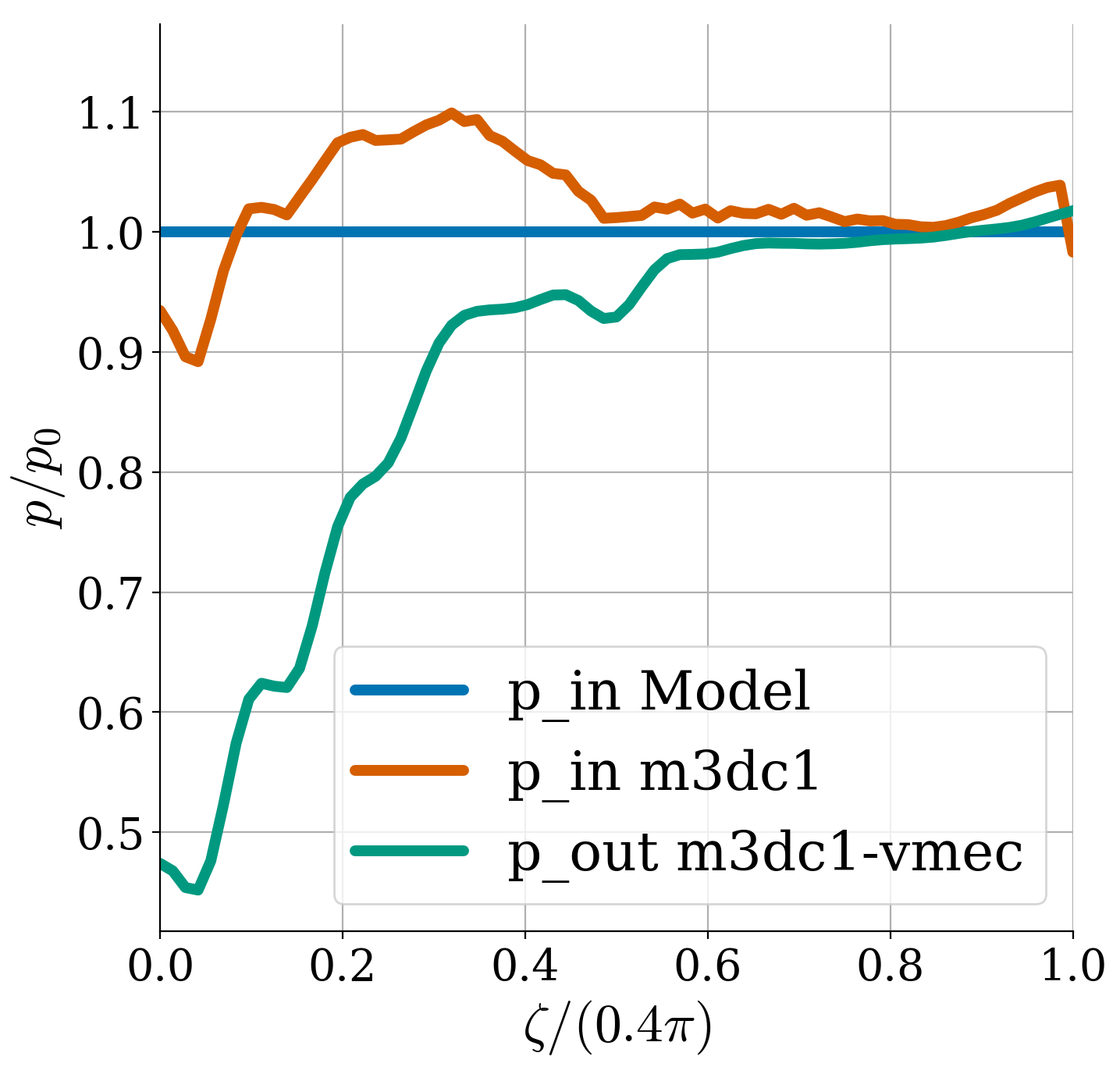}
    \caption{
        The pressure inside the flux tube is compared with pressure outside the flux tube.
        Here $p_{in}$ denotes the value of pressure in the simulation on the field line traced from the maximum of $\delta p$ at $\zeta=0$.
        The quantiy $p_{out}$ is the pressure in the VMEC equilibrium along the same field line.
        In our model, $p_{in}$ is assumed to be constant and equal to the pressure at long distances along the flux tube, 
        as indicated by the blue line.
    }
    \label{fig:W7Xdp}
\end{figure}

To further confirm that the observed structure is indeed the saturated flux tube, we 
examine two important properties of our model.
The first is the continuity of total magnetic pressure $p_{total} = p+ B^2/2\mu_0$ in the perpendicular direction
of the flux tube. As shown in Fig. \ref{fig:W7X_totalp}, the normalized total pressure perturbation
$\delta p_{total}/ p_0$ do not have the same small scale variations observed in the pressure perturbation $\delta p / p_0$ on the outboard side.
Here $\delta p_{total} = p_{total} - p_{total0}$, where $p_{total0}$ is obtained from 
the initial VMEC equilibrium, and $p_{total}$ is obtained from M3D-C1 simulation of $t=2500\tau_A$.
Therefore, we conclude that $\partial_\alpha p_{total}=0$ is a good approximation. 
The second property we check is that the pressure is constant inside the displaced flux tube. 
In Fig. \ref{fig:W7Xdp}, the pressure inside the flux tube $p_{in}$ is compared with the pressure
outside $p_{out}$. Here $p_{in}$ is the pressure on the field line traced out in the saturated stage 
of the simulation, starting at a local maximum of pressure perturbation, and $p_{out}$ is the 
pressure outside the flux tube. It is obtained by evaluating the pressure of the \textcolor{red}{initial}
VMEC equilibrium at the location of the flux tube. It represents the average pressure on that magnetic surface.
Compared with the magnitude of variation of $p_{out}$ along the flux tube, 
the magnitude of the $p_{in}$ variation is about a factor of 6 smaller. 
This means that assuming constant $p_{in}$ along the flux tube is a good approximation.

\begin{figure}[htbp]
    \centering
    \includegraphics[width=.45\textwidth]{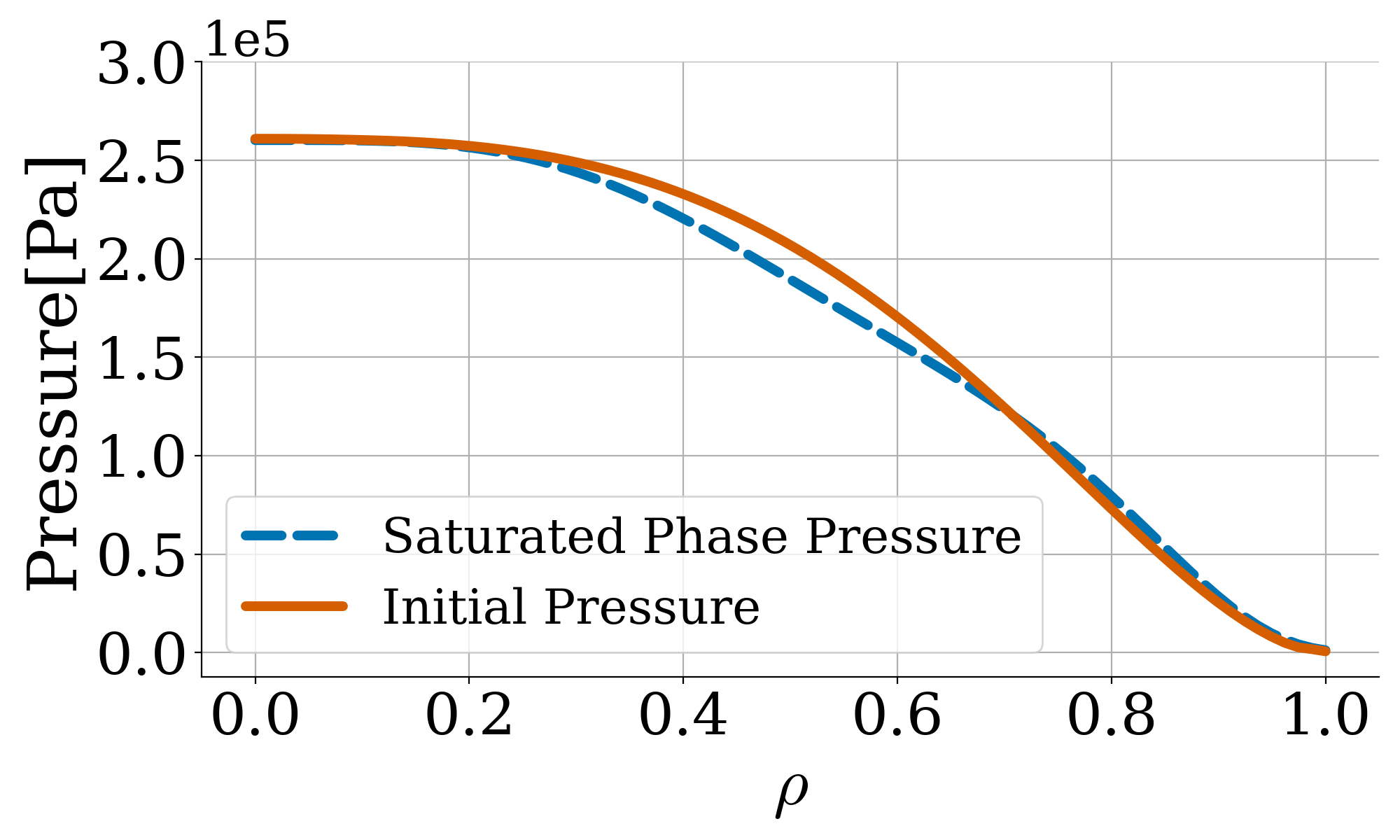}
    \caption{Average pressure profile at the saturated stage of the simulation and the initial pressure profile.}
    \label{fig:W7X_VMEC_Recon}
\end{figure}

With all the above observations, we conclude that the localized pressure perturbations observed in the simulation are 
likely to be saturated ballooning flux tube states like the ones we discuss in this paper.
To allow quantitative comparisons of the simulation with the nonlinear ballooning saturation theory, we construct 
a DESC equilibrium from the M3D-C1 results at $t=2500\tau_A$. The new pressure 
profile is obtained by surface averaging the pressure in M3D-C1 on the surfaces of the initial VMEC equilibrium.
As shown in Fig. \ref{fig:W7X_VMEC_Recon}, the pressure gradient has a slight flattening compared to the initial pressure gradient.
A new reconstructed DESC equilibrium is obtained with this new pressure profile and zero net toroidal current.

\begin{figure}[htbp]
    \centering
    \includegraphics[width=.5\textwidth]{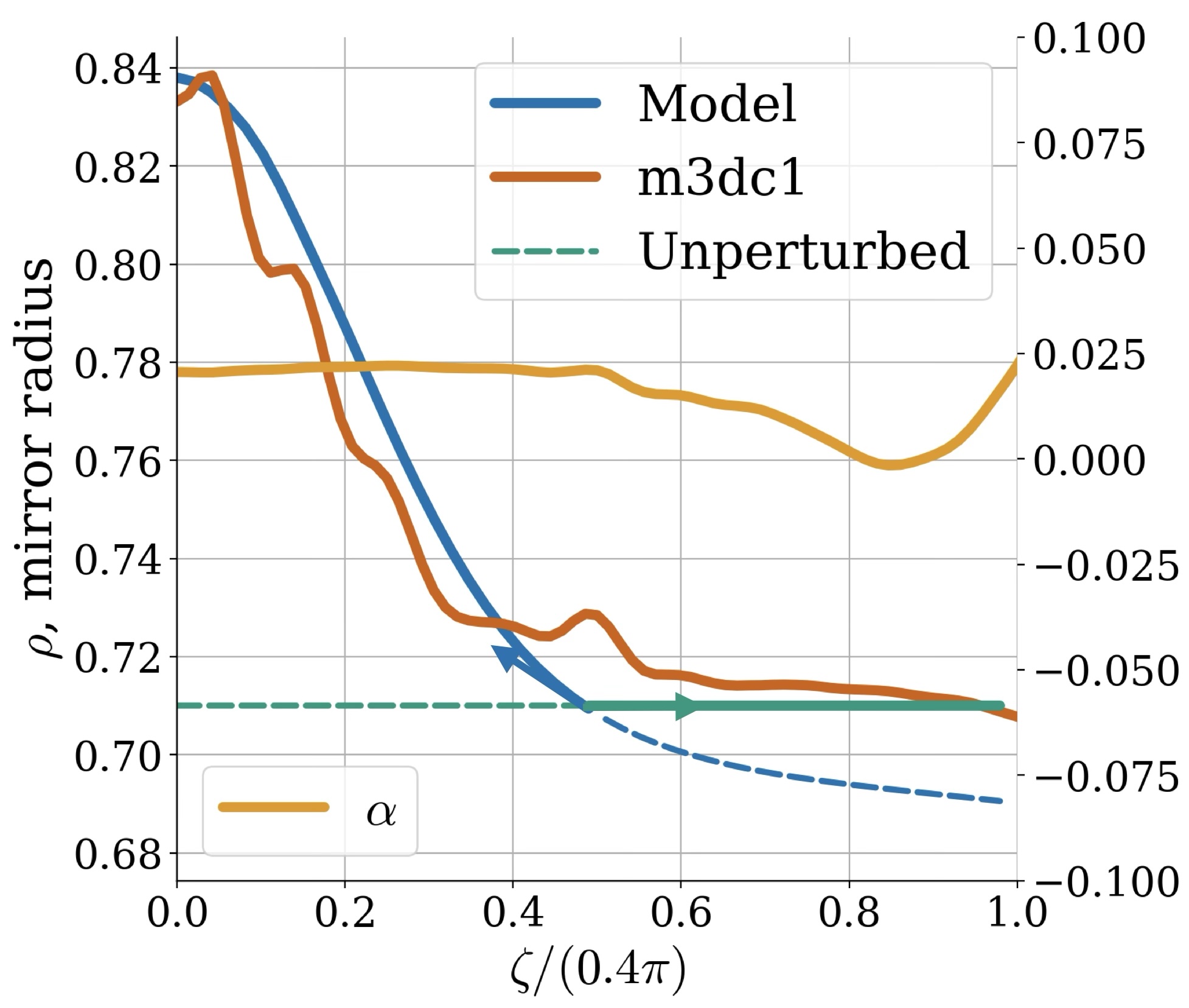}
    \caption{The flux tube shape extracted from the M3D-C1 simulation is compared with 
    a flux tube obtained using the nonlinear ballooning 
    flux tube model. The brown curve represents the flux tube extracted from the simulation by 
    tracing the magnetic field line indicated in Fig.\ref{fig:W7Xdeltap}. The blue curve 
    is the solution to the nonlinear ballooning equation with $\rho(\zeta=0)=0.837$ 
    and $\rho_0=0.71$ to match the pressure inside the flux tube.
    The yellow curve is the $\alpha_0 $ (i.e., $\theta - \iota \zeta$) value on the flux tube, with the value of $\alpha_0$ given on the axis on the right.
    }
    \label{fig:W7Xcurve}
\end{figure}

In Fig. \ref{fig:W7Xcurve}, we compare the flux tube shape extracted from the M3D-C1 simulation by following the magnetic field line 
at the maximum of $\delta p$ in $\zeta=0$ with the one calculated using the nonlinear ballooning equations Eq. \eqref{eq:ode1-fl} and Eq. \eqref{eq:ode2}. 
The flux surface that the unperturbed flux tube resides in, $\rho_0$, is prescribed so that the pressure on that surface 
equals the pressure on the flux tube extracted from the simulation at $\zeta=2\pi/5$, where
$p_{in}$ and $p_{out}$ are the same (see the blue line in Fig. \ref{fig:W7Xcurve}). 
The $\alpha$ surface is set to be $\theta - \iota \zeta=0$, which is justified 
by the alignment of $\theta - \iota \zeta=0$ surface with the pressure perturbation contour in the simulation.
The slope of the flux tube $d\eta/d\zeta$ in the model is set to zero at $\zeta = 0 $ given the stellarator symmetry of 
the configuration. As shown in Fig. \ref{fig:W7Xcurve}, the agreement between the flux tube 
shape extracted from the simulation and the flux tube calculated using our model is reasonable. 
There is good agreement in $\zeta \in (0,\pi/5)$.
The agreement is worse for $\zeta > \pi/5$, which leads to a discrepancy in the saturation level.
Since the flux tube has been obtained by choosing the value of $d\eta/\zeta$ and $\eta$ at $\zeta=0$ similar to the ones extracted from M3D-C1
instead of using the shooting method described in section \ref{sec:2},
the flux tube intersects the unperturbed flux tube at finite length along the flux tube,
and at those locations the slopes are discountinuous. 
The slope discountinuity is such that the tension of the magnetic field pulls the point radially outwards, as demonstrated by the arrows in Fig. \ref{fig:W7Xcurve}.
As a result, we would have expected the flux tube to balloon outward further.
In the simulation, zero radial displacement is 
imposed at the boundary, limiting further movement of the flux tube radially outward near the LCFS.
Comparisons with free boundary M3D-C1 simulations will be worth investigating.

\begin{figure}[htbp]
    \centering
    \includegraphics[width=.5\textwidth]{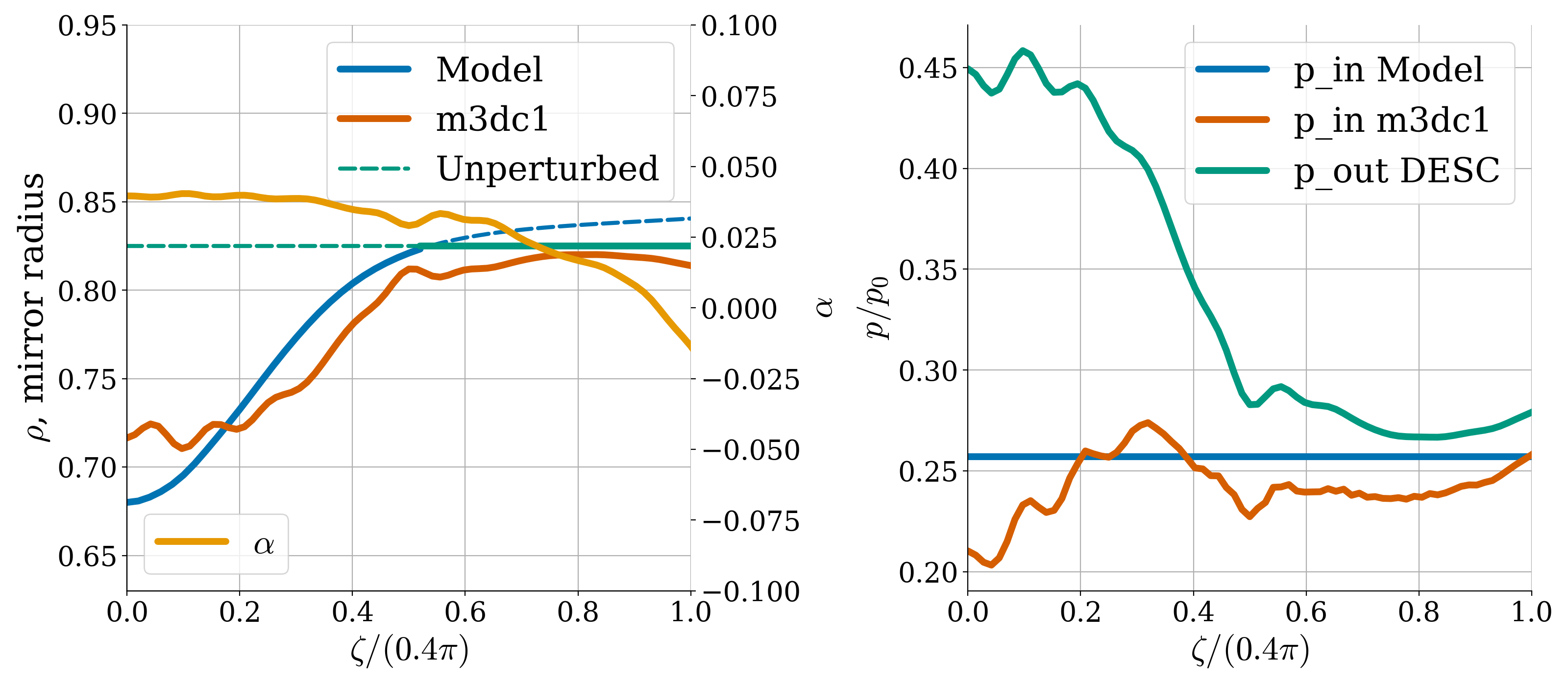}
    \caption{Inwardly ballooned flux tube. In the left panel, the brown curve represents the flux tube extracted from the simulation by 
    tracing the magnetic field line starting at the local minimum of the pressure perturbation at $\zeta=0$. The blue curve 
    is the solution to the nonlinear ballooning equation with $\rho(\zeta=0)=0.68$ 
    and $\rho_0=0.83$ to match the pressure inside the flux tube.
    In the right panel, the blue curve represents the pressure value inside the flux tube in our model.
    The brown curve is the the pressure value in M3D-C1 on the flux tube extracted from the simulation.
    The green curve is the pressure value in the DESC equilbrium on the flux tube extracted from the simulation, 
    representing $p_{out}$ in our model.
    }
    \label{fig:W7Xcurve_inward}
\end{figure}

Interestingly, we can repeat the procedure detailed above, but for a field line starting at the local minimum of the
pressure perturbation near the outboard midplane at $\zeta=0$. 
With this setup, we obtain an inwardly ballooned flux tube initially at $\rho=0.83$ and displaced to $\rho=0.72$, 
as shown in Fig. \ref{fig:W7Xcurve_inward}. Again, the constant $p_{in}$ property holds approximately.
A reasonable agreement between the flux tube shapes obtained from M3D-C1 and the nonlinear flux tube model 
with $\rho(\zeta=0)=0.68$ and $\rho_0=0.83$ is found.
    \section{Flux Tube Energy Calculation}\label{sec:4}

In this section, we describe the general behavior of the change in energy due to displaced flux tubes and the subtleties of calculating this energy change in a numerical MHD equilibrium. 
First, we show the importance of choosing the right toroidal angle to integrate the change in the energy density of the flux tube. The Boozer toroidal angle is one of the possible right choices.
Second, we show the error in the flux tube energy calculation caused 
by force errors in stellarator MHD equilibria are potentially very large. This makes 
the convergence of the energy change calculated by directly integrating Eq. \ref{eq:energy} difficult.
Finally, we introduced a variational approach to calculate the flux tube energy
that overcomes the force error problem. Convergence on equilibrium resolution is demonstrated.

\subsection{Energy Density and Choice of Toroidal Angle}
\begin{figure}[htbp]
    \centering
    \includegraphics[width=.45\textwidth]{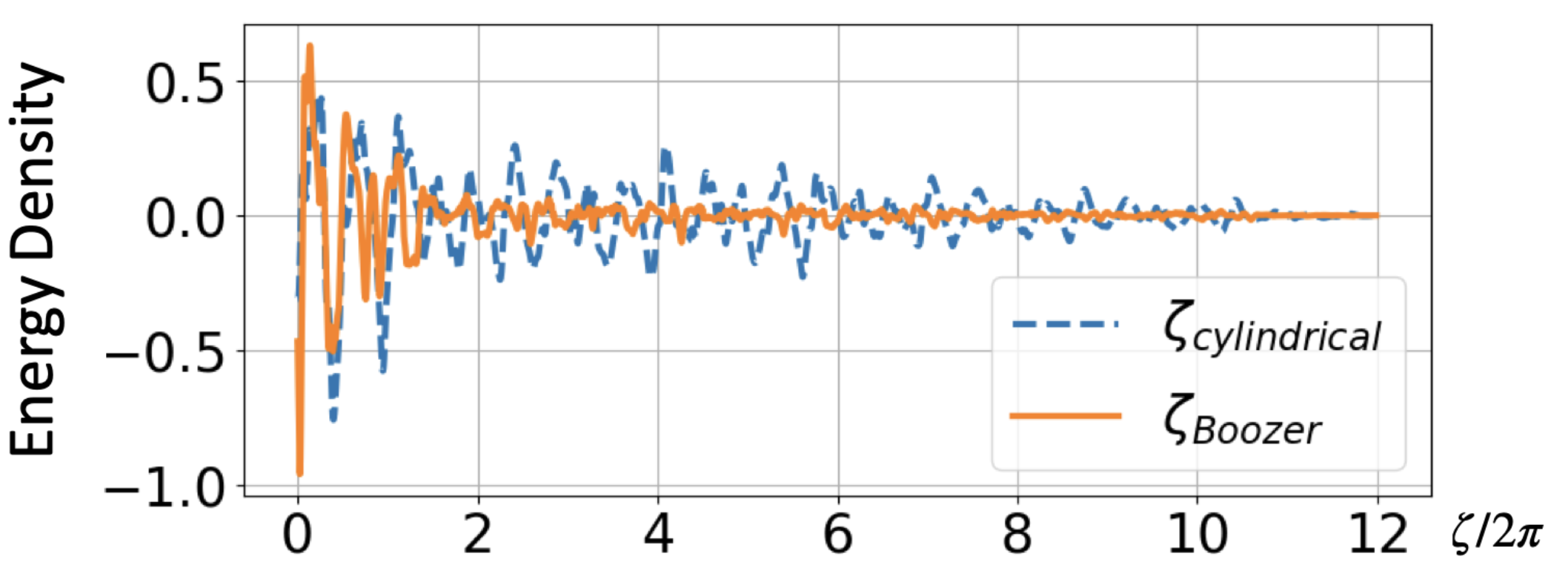}
    \caption{Energy density of a flux tube plotted against cylindrical toroidal angle versus Boozer toroidal angle}
    \label{fig:Booz}
\end{figure}

Rewriting the energy functional \eqref{eq:energy} as an integral with respect to $\zeta$, we have
\begin{align}\label{eq:energydiff}
    \mathcal{E} = \int_{-\infty}^{+\infty} \bbr{B_{in}(\rho(\zeta), \zeta)\dd{l}{\zeta} - B_0(\rho_0, \zeta) \dd{l_0}{\zeta}} d\zeta
\end{align}
where $l$ is the arc length of the field line, ${d\zeta}/{dl} = {{\bf B}_{in}\cdot\nabla\zeta}/{B_{in}}$ evaluated at $(\rho(\zeta), \zeta)$ and ${d\zeta}/{dl_0} = {{\bf B}_{0}\cdot\nabla\zeta}/{B_{0}}$ evaluated at $(\rho_0, \zeta)$.
The integrand of the energy functional (\ref{eq:energydiff}) represents the energy density of a given flux tube element. If the integrand is plotted directly as a function of 
the cylindrical toroidal angle $\zeta$, the energy has large oscillations even when the flux tube has wrapped around more than 10 toroidally turns, and the displaced flux tube approaches the unperturbed flux tube, as demonstrated by the blue dashed line in Fig.\ref{fig:Booz}. 
However, this does not mean that the part of the flux tube at large distance along its length contributes significantly to the energy. The large oscillations are due to the fact that we are not comparing with the right flux tube element of the unperturbed flux tube, as there is no reason that the perturbed flux tube element should have the same cylindrical toroidal angle as the original unperturbed flux tube element. Expanding the integrand assuming $ (\rho(\zeta)-\rho_0) \zeta\to 0$, we have
\begin{align}
    B_{in}(\rho(\zeta), \zeta)\left.\dd{l}{\zeta}\right|_{\rho(\zeta), \zeta} - B_0(\rho_0, \zeta) \left.\dd{l}{\zeta}\right|_{\rho_0, \zeta} &\sim \nonumber\\
    B_0\left.\dd{l}{\zeta}\right|_{\rho_0, \zeta, \theta = \iota_0\zeta + \iota_0'(\rho-\rho_0)\zeta} 
    -B_0\left.\dd{l}{\zeta}\right|_{\rho_0, \zeta, \theta = \iota_0 \zeta}
\end{align}
Here the dominant contribution is the difference in energy of two neighboring flux tubes located at $\theta$ and $ \theta + \iota_0'(\rho-\rho_0)\zeta$ because 
$\iota_0'(\rho-\rho_0)\zeta$ is much larger than $(\rho-\rho_0)$ when $\zeta$ is large.
If $B_0dl/d\zeta$ is dependent on $\theta$, we will have oscillations in local energy density. Note that there will not be such contribution if we are doing calculations in a tokamak and we use $\theta$ as the coordinate along the flux tube instead of $\zeta$ because by holding $\theta$ constant, we are comparing flux tube pieces that are similar.

In order to have a physical energy density such that the contribution of flux tube displacement in $\theta$ vanishes, we should use the Boozer toroidal angle $\zeta_B$, since $B_0dl/d\zeta_B=B_0^2 / \mathbf{B}_0\cd \grad \zeta_B$ is a flux function, and not dependent on $\theta$. As shown by the orange line in Fig.\ref{fig:Booz}, the energy density is now free from large oscillations at large distances along the flux tube.

\subsection{Convergence Issue due to Numerical Equilibrium Force Error}
\begin{figure}[H]
    \centering
    \includegraphics[width=.45\textwidth]{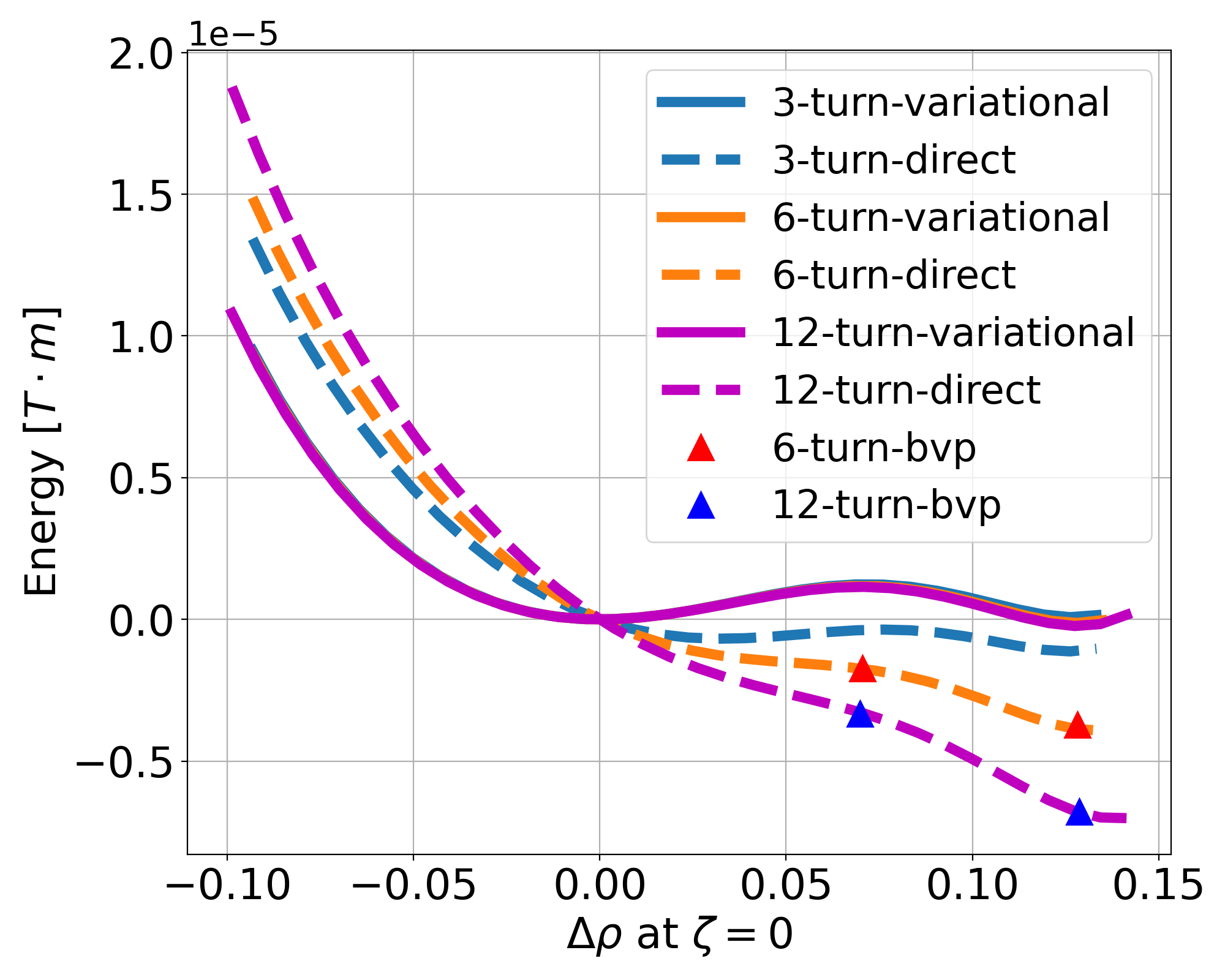}
    \caption{Comparison between the direct energy calculation and the variational method on a stellarator symmetric 
        $\alpha$ surface on the compact QA equilibrium\cite{nelson_design_2002} with $\rho_0=0.85$.
        The horizontal axis is the displacement of the flux tube at $\zeta=0$.
        Triangle markers indicate energy calculated directly for saturated flux tubes
        with different grid resolution and box size. Dashed curves shows the energy calculated directly
        for piecewise-equilibrium flux tubes with discontinuous slope at $\zeta=0$. 
        Solid curves show the energy calculated using the variational method
        for piecewise-equilibrium flux tubes with discontinuous slope at $\zeta=0$. 
        Cases with different simulation domain sizes in $\zeta$
        are indicated by different colors (one turn equals $2\pi$ in $\zeta$).
    }
    \label{fig:energy-cal}
\end{figure}

In addition to the subtlety of having to choose the variable of integration, the consistent calculation of the flux tube energy is complicated by the numerical nature of stellarator MHD equilibria. 
A direct calculation of the energy of two saturated flux tubes by direct integration of Eq. \ref{eq:energydiff} is presented as isolated triangular markers in Fig.\ref{fig:energy-cal} (at this point the reader should ignore the continuous lines which will be explained in the later part of this section). 
The result does not match the physical picture obtained from flux tube displacement calculation and linear stability analysis.
Linear ballooning stability analysis shows that the equilibrium is stable to linear ballooining modes at the minor radius of the unperturbed flux tube.
This means that the the unperturbed flux tube is on a local minimum ot the energy.
The equilibrium flux tube states with smaller displacement (corresponding to triangle markers near $\Delta \rho=0.7$ in Fig.\ref{fig:energy-cal}) should be at a locally unstable equilibrium state.
This means that the energy of the flux tube state with the smaller displacement should be larger than zero,
but the energy of this state is below zero. 
Moreover, the calculated energy of the flux tubes does not converge with the size of the domain, even though the calculated 
flux tube displacement in $\rho$ is converged.
This mismatch comes from the force error in the numerical equilibrium.
Unlike tokamak equilibria, which is proved to exist and numerically converges to zero force error, 
stellarator equilibria with integrable magnetic fields have not been proved to exist, and the numerical
equilibria are not guaranteed to converge to zero force error.

To illustrate the contribution of the force error, we compare the energy of two adjacent flux tubes by studying variations around a flux tube, 
\begin{align}\label{eq:variation}
    \delta \mathcal{E} = \int \delta \bd r \cd [ \bd B_0 \cd &\grad \bd B_0 - \bd B_{in} \cd \grad \bd B_{in} \nonumber\\
        &+ \mu_0(\grad p - \bd J \x \bd B)]\frac{dl}{B_{in}}.
\end{align}
For a detailed derivation, refer to \ref{ap:energy}.
We can apply formula \eqref{eq:variation} to the part of a saturated flux tube at large distances along the field line, 
where the displacement from the unperturbed flux tube is small and hence this variational formula gives the first order approximation to the energy contribution of this part of the flux tube.
For an equilibrium flux tube, $\delta \bd r \cd (\bd B_0 \cd \grad \bd B_0 - \bd B_{in} \cd \grad \bd B_{in}) = 0$, leading to 
\begin{align}\label{eq:varFerr}
    \delta \mathcal{E} = \int \frac{\delta \bd r \cd \bd F_{err}}{B_{in}} dl,
\end{align}
where $\bd F_{err}=\mu_0( \grad p - \bd J \x \bd B)$ is the force error of the numerical MHD equilibrium.
Expanding $\delta \bd r$ at small $\Delta \rho$, we find
\begin{align}
    \delta \bd r = \pp{\bd r}{\theta}\iota'\Delta \rho \zeta  + \pp{\bd r}{\rho} \Delta \rho
\end{align}
The leading contribution as $\zeta \to \infty$ comes from the displacement in the $\theta$ direction due to finite magnetic shear. 
As discussed in section \ref{sec:2}, the saturated flux tube radial displacement behaves as $\Delta \rho \sim \zeta^{\nu}$ when $\zeta \gg 1$.
For $-2<\nu<-1$, the integral \eqref{eq:varFerr} does not converge in general. This is supported by the behavior of energy directly calculated when changing the simulation domain size in Fig. \ref{fig:energy-cal}. 
The deviation grows with domain size while the flux tube displacement is almost unchanged.

\subsection{Overcoming the Force Error Problem: a Variational Approach}
\begin{figure}[H]
    \centering
    \includegraphics[width=.5\textwidth]{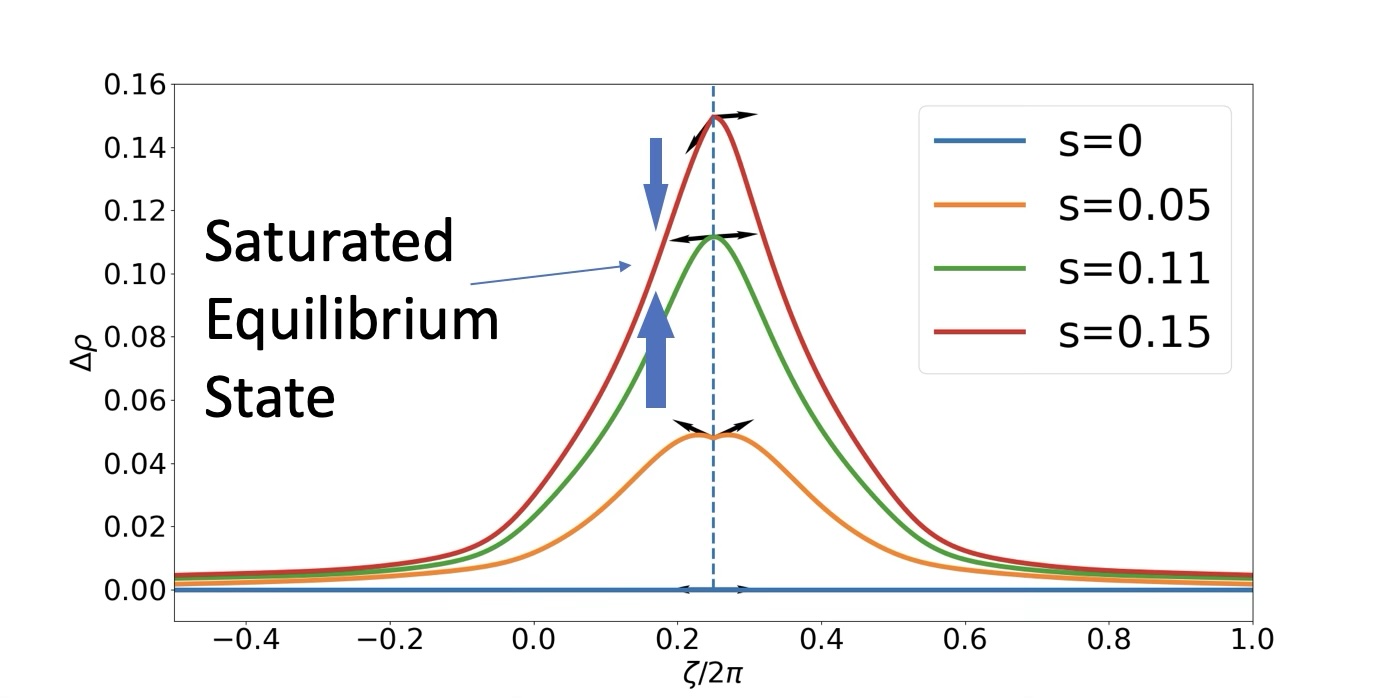}
    \caption{Shape of piece-wise $C^1$ flux tubes. 
        Here, $s$ is defined as the displacement $\Delta\rho$ at $\zeta=\zeta_{cut}$.
        Saturated states have no discontinuity in slope, as exemplified by the green curve with $s=0.11$.
        Out of equilibrium flux tubes $s=0.05$ and $s=0.15$ have a discontinuity in the slope. 
        The magnetic tension force due to the slope discontinuity points towards the saturated state.
        Calculated from a finite $\beta$ ESTELL equilibrium with a non-stellarator symmetric $\alpha$ surface and $\rho_0=0.85$.
        }
    \label{fig:pwC1}
\end{figure}

To calculate an energy that converges with box size and to be consistent with the flux tube displacement behavior, we utilize the variational formula \eqref{eq:variation}. First, to be able to perform a variational calculation, we allow the flux tube to be piece-wise $C^1$ and piece-wisely in equilibrium, that is, we allow a jump of derivative at a particular $\zeta$, $\zeta_{cut}$, as shown in Fig. \ref{fig:pwC1}. 
The sections on either side of the flux tube are in equilibrium, i.e. they satisfy Eqs. \eqref{eq:ode1-fl} and \eqref{eq:ode2}. With this discontinuity in 
slope, there is now a continuous family of piece-wise equilibrium flux tubes, labelled by the displacement $s=\Delta\rho_{cut}$ at $\zeta_{cut}$.
The flux tubes with continuous slope in this family are the flux tubes in force balance.

We use an updated version of the variational formula that takes discontinuity of slope into account (derived in \ref{ap:energy}, Eq. \eqref{eq:ap:discounty}),
\begin{align}
    \delta \mathcal{E} = \int_{-\infty}^{+\infty} \frac{1}{B_{in}} &\bbr{\bd B_0 \cdot \grad \bd B_0 - \bd B_{in} \cdot \grad \bd B_{in} + \bd F_{err}} \cdot \delta \bd r dl 
    \nonumber\\- &\bb{{\bd B}_{in} }_{cut}\cd \delta \bd r_{cut}.
\end{align}
Here $[A]$ denotes the jump of quantity $A$ across $\zeta=\zeta_{cut}$.
The variation $\delta \bf r$ is a continuous vector function of $l$.
The contribution from the force error (the first term) and the rest (the second term) can now be separated. The part that is independent of force error can be calculated by 
integrating $\bb{{\bd B}_{in}}\cdot \frac{\delta \bd r_{cut}}{\delta s}$ in displacement $s$. It can be readily seen that the energy calculated in this way is consistent with the behavior of 
flux tube displacement: since $\bb{{\bd B}_{in}}$ is zero for equilibrium flux tubes, they are automatically the stationary point of the energy curve. With this method,
the procedure for calculating the energy of an equilibrium flux tube is as follows. First, a family of flux tubes is obtained by integrating from the two ends of the simulation
domain to where the cut is. All the flux tubes have zero radial displacement at the ends of the domain by construction. Each flux tube ends at a different radial location at the $\zeta$ cut. Note that it is not necessary that flux tubes calculated in this way join at the cut
unless the $\zeta_{cut}$ is chosen to be at a symmetry point.
With the family of flux tubes calculated in this way, $\bd B_{in}$ on both sides of the cut can be interpolated as a function of radial displacement at the cut, and then $[\bd B_{in}]$ can be 
integrated to calculate the energy of each piece-wise equilibrium flux tube. The stationary point of the energy curve correspond to the energy and displacement of 
the equilibrium flux tubes. We can also get the shape of the equilibrium flux tubes by interpolating the family of the piece-wise equilibrium flux tubes as a 2D function
$\Delta \rho(\zeta, s)$, where the radial displacement at the cut $s$ works as a flux tube label. 
The equilibrium flux tubes shape are $\Delta \rho(\zeta, s_{eq})$, where $s_{eq}$ are the radial displacement at the cut that correspond to a stationary point on the energy curve.

This method is effectively a nonlinear extension of Newcomb's principle\cite{newcomb_hydromagnetic_1960}. 
In the linear limit, if the solution crosses zero, we can truncate the solution at the intersection and join it with the unperturbed flux tube. Using 
the argument of field line tension described in Fig.\ref{fig:pwC1}, the flux tube tends to move upwards at the cross-over location, i.e. it is unstable.

As a side note,
using the shooting method to find the saturated flux tubes, described in section \ref{sec:2}, requires first mapping out the flux tube behavior by varying the slope at one end. 
This generates a curve of $\Delta \rho (\zeta=\zeta_1)$ v.s. $Y(\zeta_0)$. The interval to find the correct $Y$ is the one that includes the value of $Y(\zeta_0)$ where $\Delta \rho(\zeta=\zeta_1)$ crosses zero.
Once the bounds of the slope are obtained, a root finding algorithm is used. 
Compared with the variational approach, the shooting method requires a coarser scan of the slopes, but the second root finding step requires additional ODE IVP solves and it is serial. 
Instead, the variational approach can be easily parallelized. It also generates the energy value using the discontinuity in slope at the cut. 
The variational approach is typically more computational advantageous.

We plot the energy of a family of flux tubes as a function of $s$ in Fig. \ref{fig:energy-cal}.
The energy curves are insensitive to the simulation domain size. The dashed lines are the energy curves calculated for the family of piece-wise equilibrium flux tubes using 
Eq. \eqref{eq:energydiff}. The difference between the two sets of curves is the contribution from the equilibrium force error. This contribution is significant, especially near marginality where the energy is small.

\begin{figure}[H]
    \centering
    \includegraphics[width=.5\textwidth]{"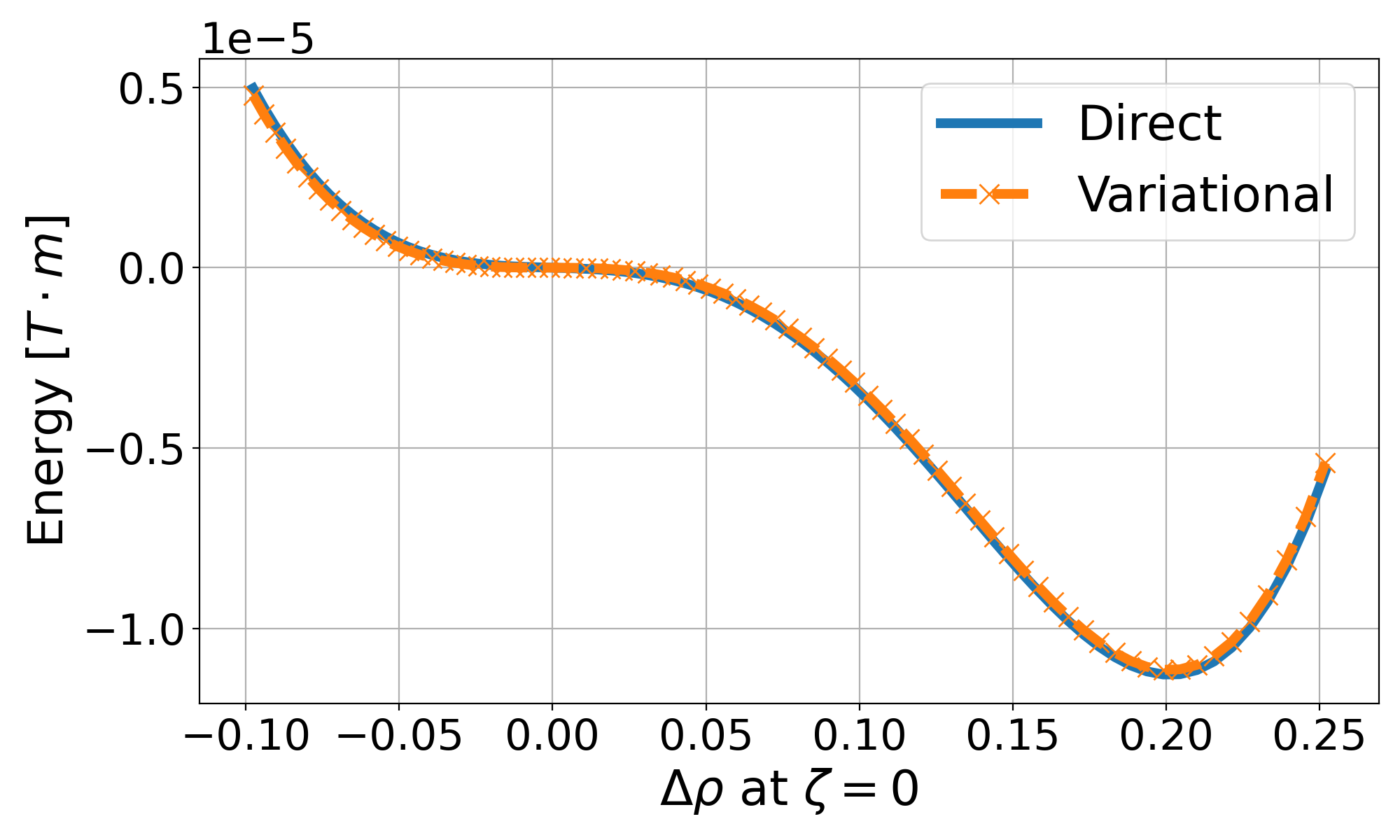"}
    \caption{Comparison between the piecewise $C^1$ flux tube energy 
        calculated using the variational method and the same energy calculated using the direct method in a tokamak equilibrium 
        (
        The DSHAPE Case in \cite{panici_desc_2023}
        ). 
        The $\alpha$ surface is up-down symmetric on the outboard mid-plane.
        The simulation domain is $\zeta \in [0,24\pi]$.
        The relative force error $F_{error}/|\grad B^2|$ is smaller than $10^{-6}$ in this equilibrium.
        }
    \label{fig:energy-curve-DTok} 
\end{figure}

To further illustrate the problem with the equilibrium force error, 
we confirm in Fig. \ref{fig:energy-curve-DTok} that in a tokamak equilibrium, where the force error is three orders of magnitude lower than that in the compact QA case\cite{nelson_design_2002} that we are using, the two ways of calculating the energy match each other.

\begin{figure}[H]
    \centering
    \includegraphics[width=.5\textwidth]{"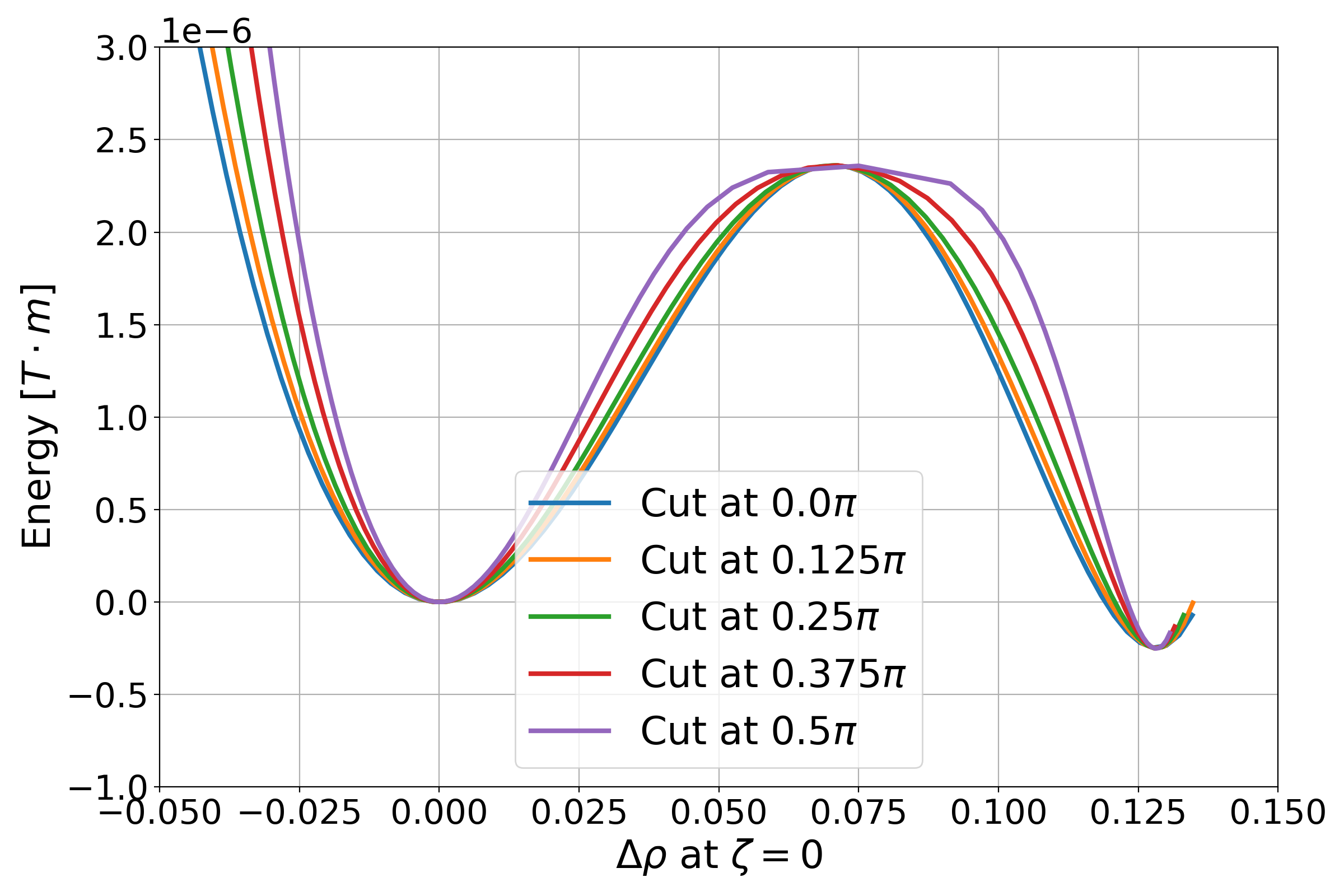"}
    \caption{
        Energy of nonlinearly saturated flux tubes in a compact QA equilibrium\cite{nelson_design_2002} calculated using the variational method with different location of 
        the slope discontinuity $\zeta_{cut}$. The horizontal axis is the displacement
        of the flux tubes at $\zeta=0$.
        The local maximum and minimum of energy curve correspond to the saturated flux 
        tube state. They remain unchanged when $\zeta_{cut}$ varies.
    }
    \label{fig:energy-curve-cut} 
\end{figure}

Given that the cut location is arbitrary, the energy of the saturated flux tubes should be independent of this choice. 
This is demonstrated by Fig. \ref{fig:energy-curve-cut}, which illustrates that the stationary points of all energy curves generated with
differing cut locations coincide. 

\begin{figure}[H]
    \centering
    \includegraphics[width=.5\textwidth]{"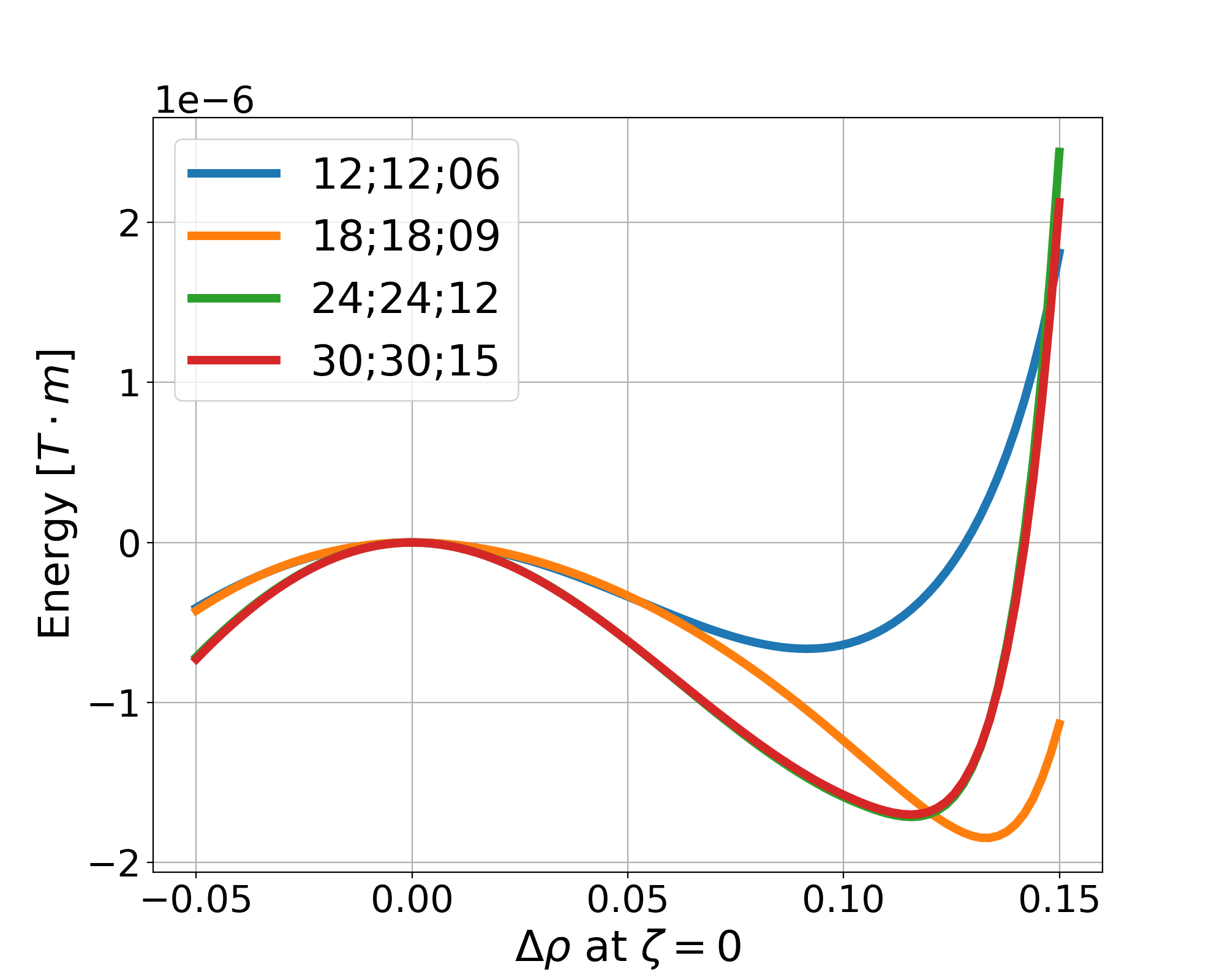"}
    \caption{
        Convergence of flux tube energy curves, calculated using the variational method, for the same 
        unperturbed flux tube with different equilibrium resolution.
        The equilibrium resolution is labelled by a tuple of three integers, each representing radial mode number, poloidal mode number and toroidal mode number.
        The equilibrium is for a finite beta version of ESTELL at $\rho_0=0.85$.
    }
    \label{fig:energy-curve-conv}
\end{figure}

We have found that the nonlinear ballooning saturation calculation is convergent
with respect to numerical choice in the solver and \textcolor{red}{box size}. 
In Fig. \ref{fig:energy-curve-conv} we demonstrate the convergence with respect to numerical equilibrium resolution.
We generated four equilibria with increasing resolution for an ESTELL fixed boundary equilibrium\cite{drevlak_estell_2013}.
Due to the nature of the variational calculation method, the energy of a flux tube converges as long 
as the flux tube shape calculation converges.

One thing to note is that,  similar to linear ballooning calculations,
the nonlinear ballooning calculation is sensitive to 
small changes in equilibrium. A correct calculation usually requires a high resolution
for the numerical equilibrium and a careful convergence analysis. As a rule of thumb, 
the nonlinear results is usually correct when the linear ballooning growth rate converges
in the entire domain of the nonlinear calculation (which also indicates convergence of the Mercier stability calculation).

We have found stellarator equilibria
whose nonlinear ballooning calculation does not converge with increasing equilibria resolution to the level of resolution that we could afford. 
However, in those cases, their linear ballooning growth rate did not converge either.
The study of this lack of convergence is out of the scope of this paper.
    \section{Case study on a compact QA\cite{nelson_design_2002} equilibrium}\label{sec:5}

In this section we study the behavior of nonlinear saturated flux tubes
in compact QA equilibria\cite{nelson_design_2002}. First we will show that in the standard 
configuration of \cite{nelson_design_2002}, stable, unstable and metastable magnetic flux tubes exit.
Then, we construct a modified equilibrium with slightly lower 
$\beta$, just below the linear instability boundary, and we show that in this 
globally linearly stable case, metastable flux tubes exist.

We follow the following procedure for each equilibrium. For each equilibrium configuration, 
linear stability is analyzed near the stellarator last closed flux surface(LCFS). Then, for each radial 
location, the nonlinear ballooning equation is solved on the stellarator symmetric $\alpha$ surface
centered at $\theta=0,\,\zeta=0$ to find saturated flux tubes that do not balloon outside the LCFS. 
For each saturated flux tube, we also calculate the energy (Eq. \eqref{eq:energy}).
We have chosen the $\alpha$ surface to be stellarator symmetric because in most flux surfaces, the most unstable 
modes are symmetric with respect to $\zeta=0,\theta=0$.

\subsection{Standard Configruation with Linear Ballooning Instability}

\begin{figure}[H]
    \centering
    \includegraphics[width=.5\textwidth]{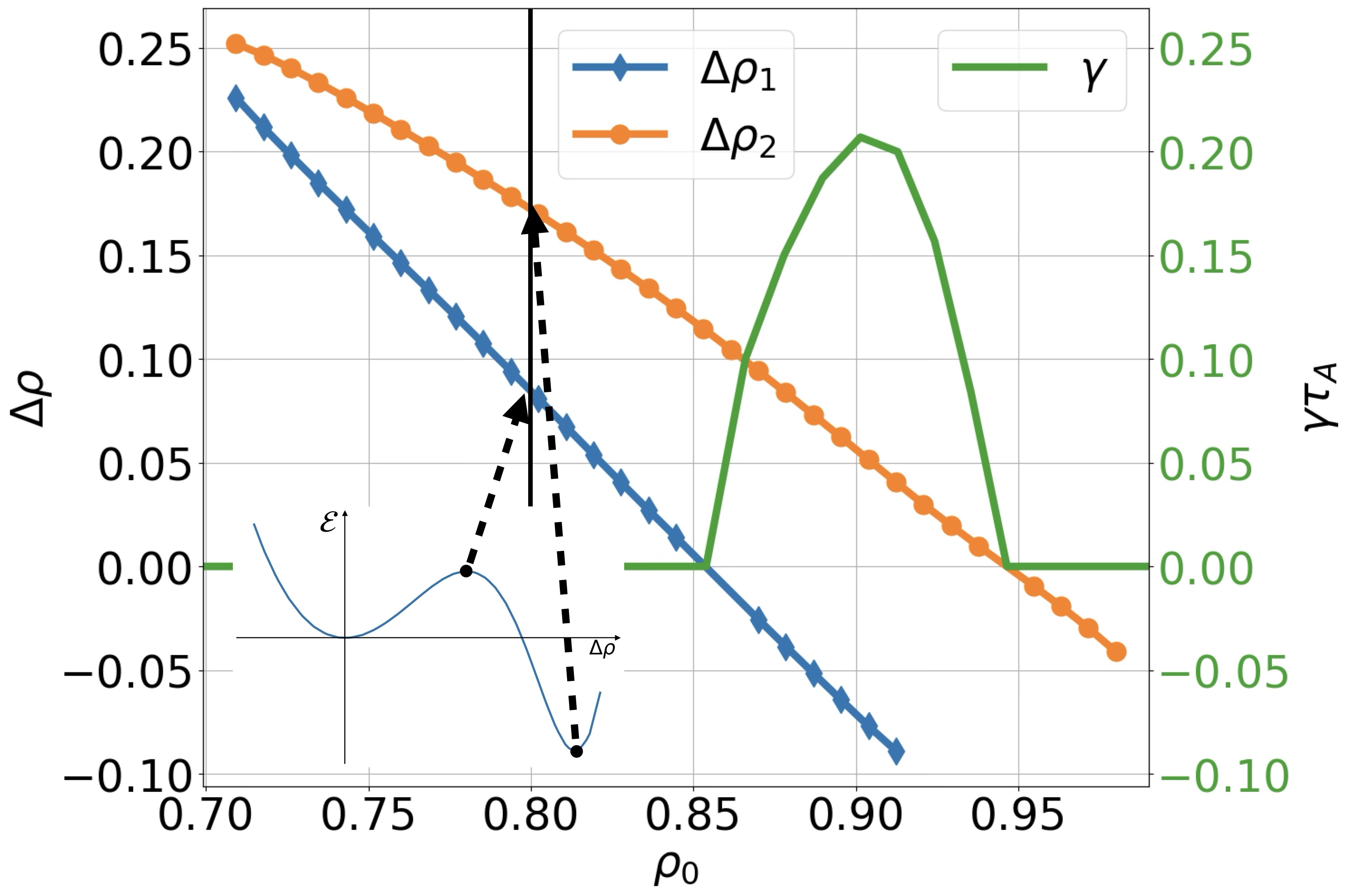}
    \caption{Saturated flux tube maximum displacement calculated on a compact QA\cite{nelson_design_2002} equilibrium.
        Here, $\Delta \rho$ denotes the displacement of the flux tube at $\zeta=0$, and 
        $\rho_0$ denotes the flux surface label of the original, unperturbed flux tube.
        We also plot the normalized (to Alfven frequency) 
        linear ballooning growth rate $\gamma$ (values given on the right).
        The corresponding energy curve of the flux tube initially at $\rho=0.8$ is shown on the 
        lower left corner. The two branches correspond two the stationary points of the energy curve as
        indicated.
    }
    \label{fig:NCSXdisp1}
\end{figure}

\begin{figure}[htbp]
    \centering
    \includegraphics[width=.5\textwidth]{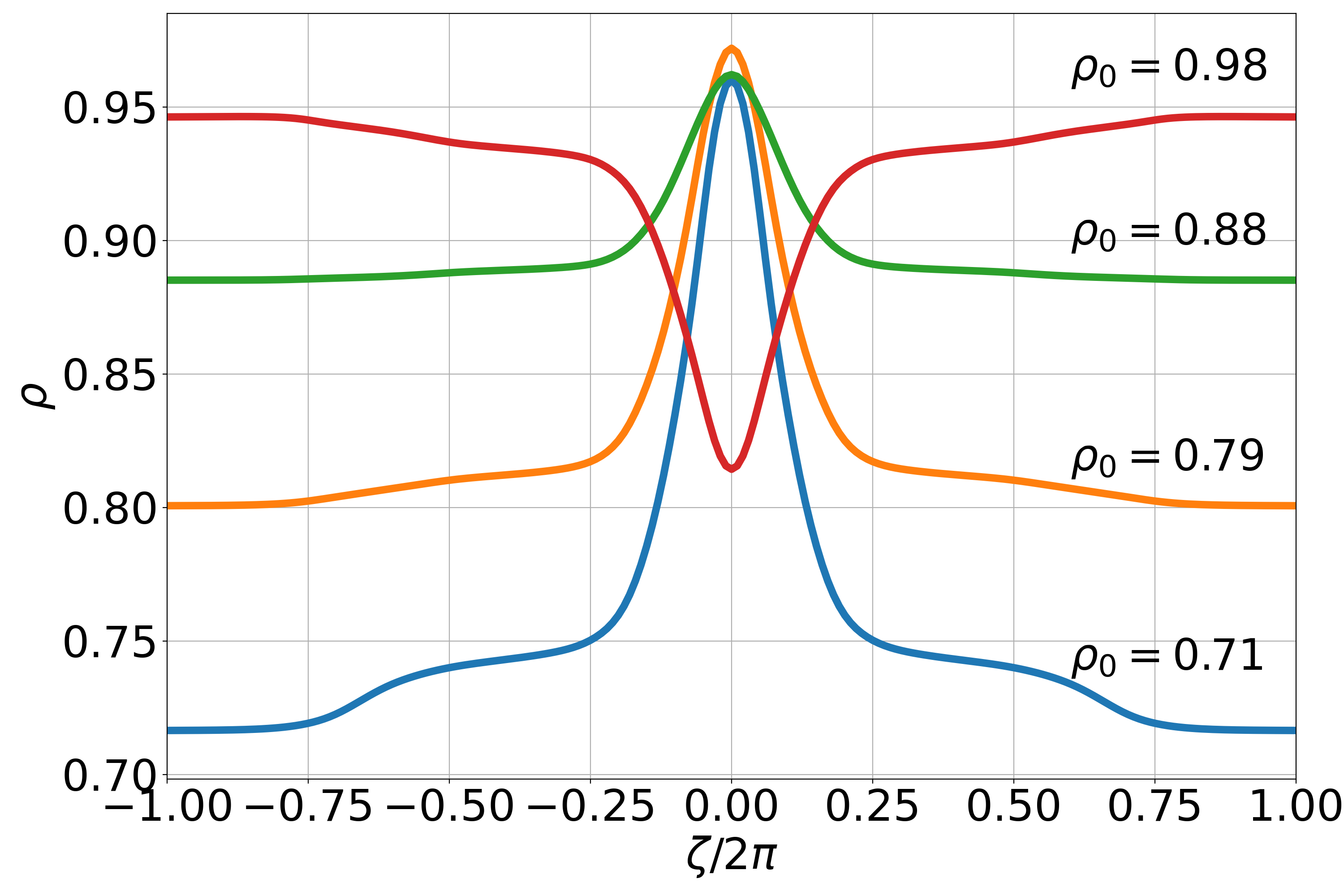}
    \caption{Saturated flux tube shape on the compact QA equilibrium\cite{nelson_design_2002}.
        The \textit{y} axis is the minor radius $\rho$ instead of displacement $\Delta \rho$. 
        Note that for the $\rho_0=0.98$ case, the $\Delta \rho_1$ branch of Fig. \ref{fig:NCSXdisp1} is plotted 
        since it has a lower energy. 
        For the rest of the cases, the $\Delta \rho_2$ branch has lower energy and are plotted.}
    \label{fig:NCSXline1}
\end{figure}

In Fig. \ref{fig:NCSXdisp1}, the maximum radial displacement along the saturated flux tube is plotted
as a function of the radial location of the corresponding unperturbed flux tube, for the compact QA\cite{nelson_design_2002} standard configuration.
There are two different branches of saturated flux tubes, each corresponding to a 
stationary point on the energy curve, as indicated by the energy curve in the subfigure in Fig. \ref{fig:NCSXdisp1}.
On top of the displacement, the linear ballooning growth rate is also plotted for reference.
As indicated, there is a linear unstable region from minor radius $\rho=0.85$ to $0.95$. 
Some typical saturated flux tubes of the second branch (orange colored) are plotted 
in Fig. \ref{fig:NCSXline1}. 
Note that the $y$ axis in Fig. \ref{fig:NCSXline1} is the minor radius $\rho$ of instead of the radial displacement $\Delta \rho$ and 
therefore different saturated flux tubes go back to different $\rho_0$, their unperturbed location.


\begin{figure}[H]
    \centering
    \includegraphics[width=.5\textwidth]{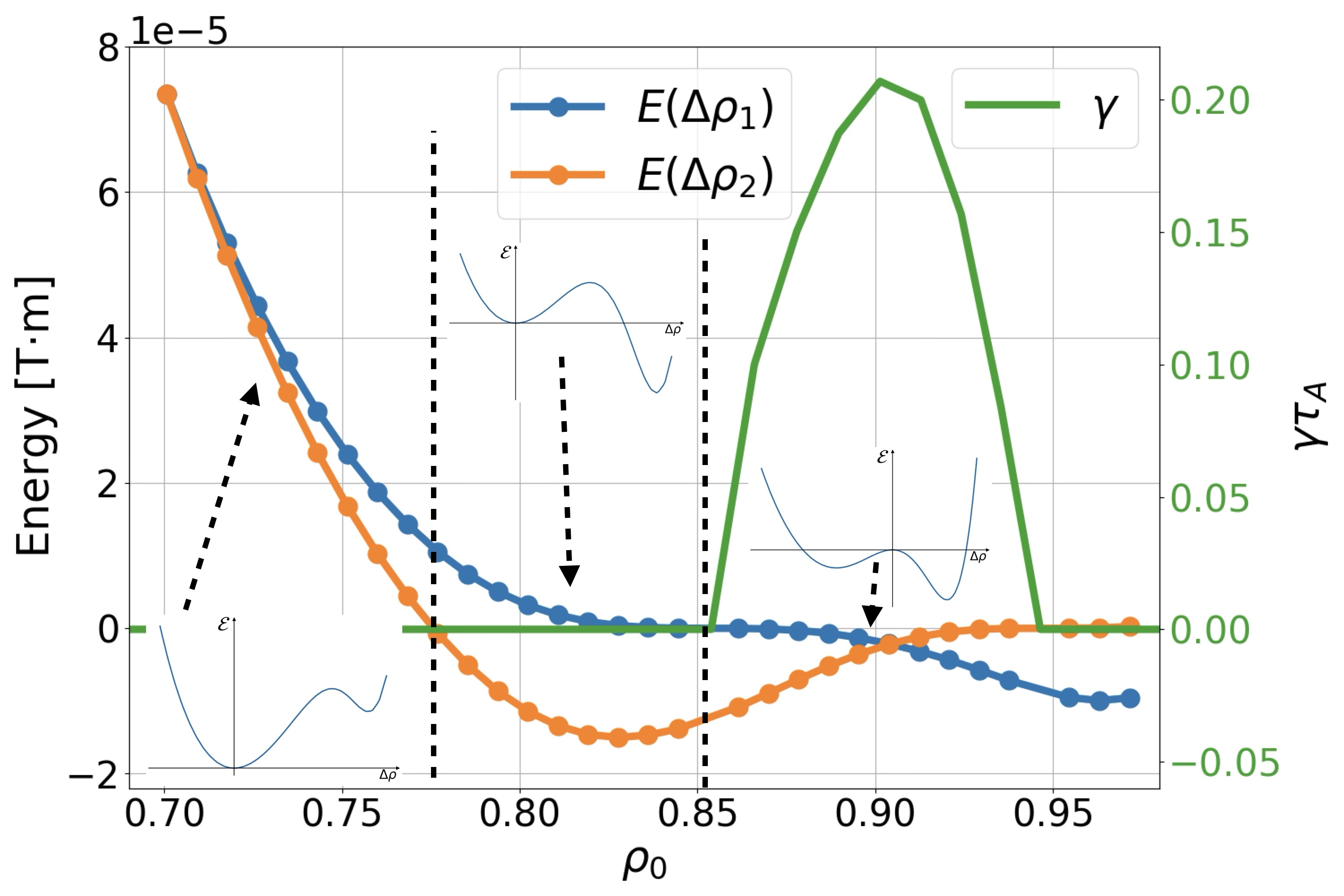}
    \caption{Saturated flux tube energy calculated on the compact QA\cite{nelson_design_2002} equilibrium.
    Here, $\rho_0$ denotes the flux surface label of the original, unperturbed flux tube.
    We also plot the normalized linear ballooning growth rate $\gamma$ (see values on the right). 
    }
    \label{fig:NCSXen1}
\end{figure}

In Fig. \ref{fig:NCSXen1}, the energy of the flux tubes in Fig. \ref{fig:NCSXdisp1} is plotted.
As we can see, the nonlinear saturated flux tubes exist from minor radius $\rho_0=0.7$ all the way to the LCFS.
For $\rho_0\in (0.7,0.77)$, both branches of the saturated flux tubes $\Delta\rho_1$ and $\Delta \rho_2$ have higher
energy then the original flux tube, making the plasma nonlinearly stable to localized perturbations in this
region. For $\rho_0 \in (0.77,0.85)$, the branch $\Delta \rho_2$ of the saturated flux tubes has lower energy then the original flux tube, 
whereas the branch $\Delta \rho_1$ still has higher energy then the original flux tube. 
In this region, flux tubes are metastable, with a maximum 
displacement of 15\% of the minor radius. In both of these regions, the saturated flux tubes in 
branch $\Delta \rho_1$ correspond to local maxima of the energy curve. The flux tubes on the branch $\Delta \rho_1$ are hence unstable and in reality only
branch $\Delta \rho_2$ of the saturated flux tubes or the unperturbed flux tubes can exist.
For $\rho_0 \in (0.85,0.95)$, the ballooning mode becomes linearly unstable as indicated by the non-zero linear 
growth rate. Both branches of the perturbed flux tubes have lower energy and the nonlinear theory predicts
their saturation levels. One of the branches balloons radially inwards, and the other branch balloons radially 
outwards, which is expected since the leading order term of the energy functional is quadratic in amplitude.
In the region close to marginality, one of the two branches of the saturated flux tube should 
have small saturated level and small energy release, and go to zero at the marginally stable surface.
This is indeed the case in the nonlinear calculation, where the saturated flux tube displacement and 
energy curves go to zero exactly where linear growth rate goes to zero.

In the compact QA\cite{nelson_design_2002} standard configuration, nonlinearly stable, metastable and linearly unstable flux tubes exist in different minor radii. 

\subsection{Modified Linearly Stable Configuration: Metastabilty}
\begin{figure}[H]
    \centering
    \includegraphics[width=.5\textwidth]{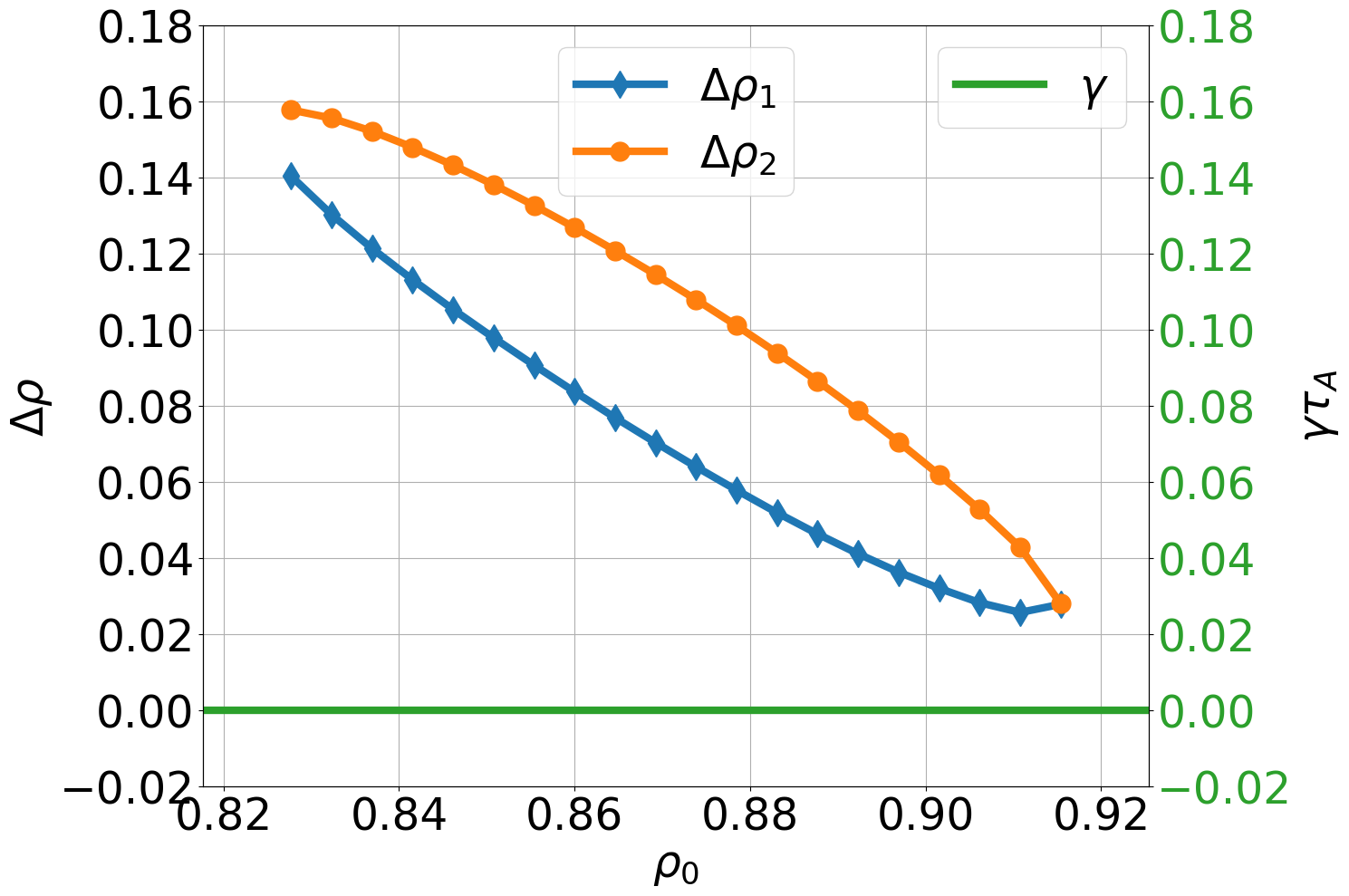}
    \caption{Saturated flux tube maximum displacement calculated in the compact QA equilibrium\cite{nelson_design_2002} with 
    scaled pressure profile.
    Here, $\Delta \rho$ denotes the displacement of the flux tube at $\zeta=0$,
    $\rho_0$ denotes the flux surface label of the original, unperturbed flux tube.
    We also plot the normalized linear ballooning growth rate $\gamma$ (see values on the right), 
    indicating that this $\alpha$ surface is globally stable.}
    \label{fig:NCSXdisp2}
\end{figure}

\begin{figure}[H]
    \centering
    \includegraphics[width=.5\textwidth]{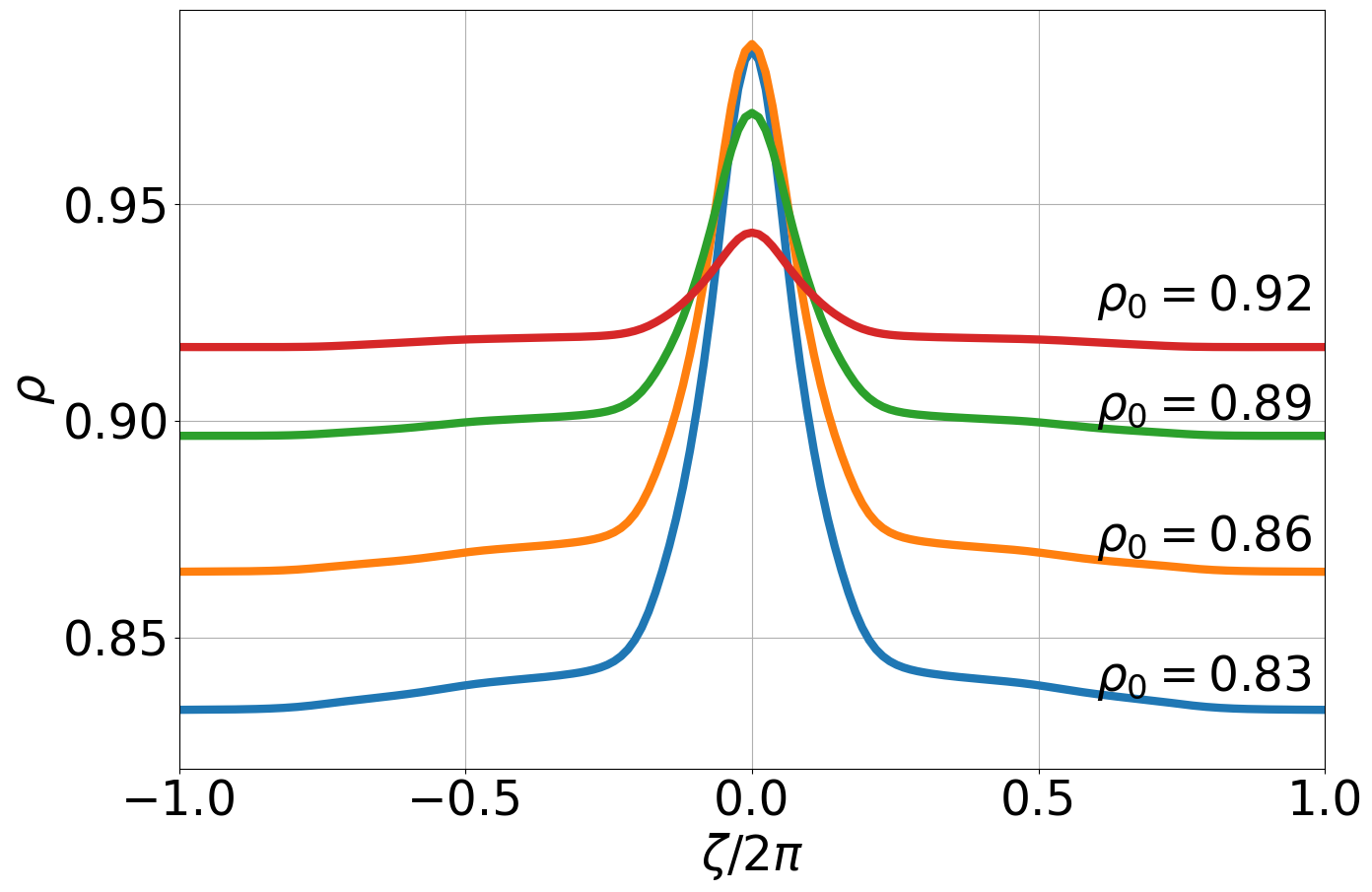}
    \caption{Saturated flux tube shape on the compact QA equilibrium\cite{nelson_design_2002} for branch $\Delta \rho_2$ in Fig. \ref{fig:NCSXdisp2}.
    The \textit{y} axis is the minor radius $\rho$ instead of displacement $\Delta \rho$.}
    \label{fig:NCSXline2}
\end{figure}

\begin{figure}[H]
    \centering
    \includegraphics[width=.5\textwidth]{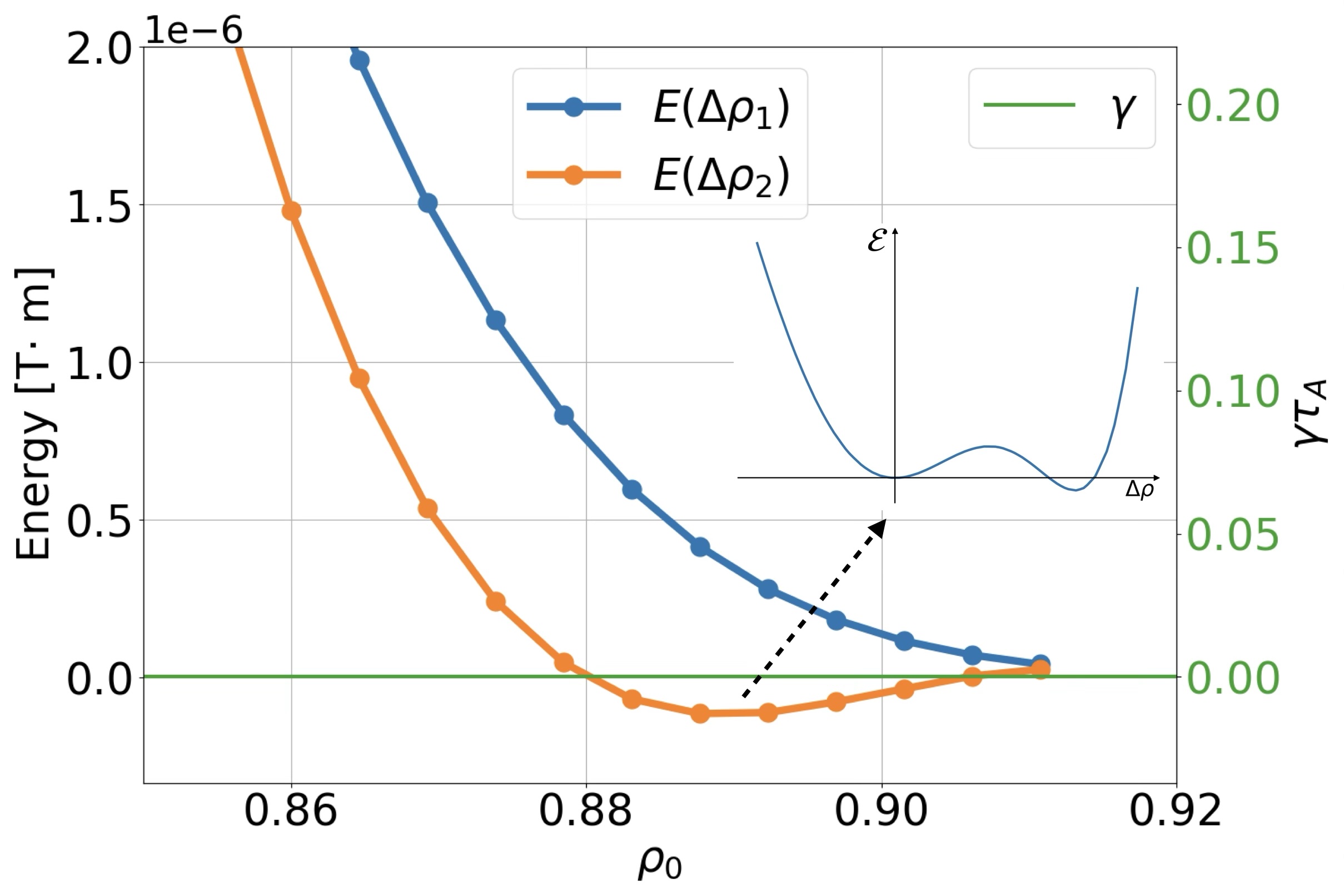}
    \caption{
        Saturated flux tube energy calculated on the compact QA\cite{nelson_design_2002} equilibrium.
        Here, $\rho_0$ denotes the flux surface label of the original, unperturbed flux tube. We also plot
        the normalized linear ballooning growth rate $\gamma$ (values on the right).
        The arrow points to the energy curve for the piecewise $C^1$ flux tubes 
        calculated for $\rho_0=0.885$, showing the existence of metastable state.
        }
    \label{fig:NCSXen2}
\end{figure}

We scale down the pressure profile in the standard configuration by \textcolor{red}{3\%} so that the new configuration is linearly stable to ballooning modes everywhere(at worst marginally
stable) and perform the same nonlinear ballooning analysis procedure used above.
The nonlinear saturated flux tube displacement and energy are plotted in Figs. \ref{fig:NCSXdisp2}, \ref{fig:NCSXen2} and 
\ref{fig:NCSXline2}.


At $\rho_0=0.89$, we can see that despite the equilibrium being linearly stable 
across the entire minor radius, there still exist metastable flux tubes.
Note that the energy release is small compared to the metastable flux tubes 
in the standard compact QA equilibrium\cite{nelson_design_2002}. However, this is expected to be sensitive to how close 
the equilibrium is to marginality and the radial extent of the marginally 
stable region\cite{ham_nonlinear_2018}. 

    \section{Conclusions}\label{sec:7}

We have developed a set of novel tools to calculate the nonlinear saturation of ballooning modes in stellarators using the model in \cite{ham_nonlinear_2018}
and the DESC equilibrium solver\cite{panici_desc_2023}.
Convergence analysis and benchmarks against linear ballooning codes have been performed.
A variational method for flux tube energy calculation was developed to eliminate the contribution 
due to the finite force error in the stellarator numerical equilibrium.
This contribution impeded the convergence of the direct way of calculating 
energy.

By analyzing a recent simulation of W7X using M3D-C1\cite{zhou_benign_2024},
we show that nonlinear saturated flux tube structures like the ones proposed in \cite{ham_nonlinear_2018} and in this article
exist in a global nonlinear MHD simulation. 
In the nonlinear saturated stage, the pressure perturbation 
has localized peaks in a toroidal cross-section. These peaks of the pressure perturbation roughly follow field lines, thus resembling 
the flux tube in \cite{ham_nonlinear_2018}. 
The shape of the pressure perturbation contour near the peak is elongated along an $\alpha$ surface.
Total pressure balance across the $\alpha$ surface and hydrodynamic pressure
balance along the flux tube were each checked to hold approximately. We have compared the flux tube displacement extracted from 
the simulation with our nonlinear ballooning model, finding good agreement. 

A study of modified compact QA equilibria\cite{nelson_design_2002} shows the existence of stable, unstable and metastable flux tubes. 
When the equilibrium is linearly stable globally but close to marginal stability, lower energy states 
can also be found, thus the unperturbed equilibrium is metastable. This indicates that explosive MHD behavior can also happen in stellarator.
The existence of such metastable flux tube states might be related to ELM-like bursts observed in stellarators\cite{hirsch_major_2008}.

One thing to note here is that the calculation of the nonlinear saturated state is very sensitive to 
the numerical equilibrium. This behavior is also observed in linear ballooning growth rate calculations.
Careful convergence studies of the equilibrium resolutions need to be conducted for each calculation. 
The results might not converge for some highly shape equilibria, for which the required resolution 
is not attainable with the computation power that we had. A general rule of thumb is that the nonlinear ballooning calculation
is convergent as long as the linear ballooning growth rate calculation has converged for each field line on each flux 
surface. Further studies on which part of the equilibrium 
lead to this sensitivity and how to reduce such sensitivity are needed.

Compared with tokamaks, no qualitative difference in the existence and behavior of nonlinear saturated flux tube states has been
observed so far. Near marginal stability, the second order terms in the energy principle approach zero. As long 
as there is a non-zero third order term, metastable equilibria generally exist. This can be backed up by 
the fact that we found metastable equilibria in our case study and no special changes of the 
magnetic geometry were introduced to find them. Lower energy saturated states even exist in configurations that are linearly stable over the whole volume.
The energy release is small in the metastable cases we presented, but this 
is expected to be configuration dependent. From empirical experience with this model on tokamaks, increasing 
the radial extent of marginal stable region generally increases the energy release of the metastable flux tubes\cite{ham_nonlinear_2018}.
    \section*{Acknowledgements}
The authors would like to thank Dario Panici and Rahul Gaur for the DESC code support.
This work is supported by the U.S. Department of Energy under contract number DE-AC02-09CH11466.
Y.Z. is supported by the National MCF R\&D Program under Grant Number 2024YFE03230400 and the Fundamental Research Funds for the Central Universities.
The United States Government retains a non-exclusive, paid-up, irrevocable, 
world-wide license to publish or reproduce the published form of this manuscript,
or allow others to do so, for United States Government purposes.

    \renewcommand\refname{References}
    \providecommand{\newblock}{}

\appendix
\section{Numerical implementation of the nonlinear ballooning equations \eqref{eq:ode1-fl} and \eqref{eq:ode2}}\label{ap:nu}

Initially, the computational domain, namely the $\alpha$ surface, is set up.
We construct a 2D uniform grid in $\zeta_{DESC}$ and $\rho$ on the $\alpha$ surface.
The alpha surface is defined by
\begin{align}
    \alpha = \theta_{DESC} + \lambda_{DESC} - \iota \zeta_{DESC} - f(\rho),
\end{align}
where $f$ is a prescribed function, such as a polynomial. Here, $\zeta_{DESC}$ is the toroidal angle used in DESC, which is 
the same as the cylindrical toroidal angle, $\theta_{DESC}$ is the poloidal angle used in DESC, and $\lambda_{DESC}$ is the difference 
between the straight field line poloidal angle corresponding to $\zeta_{DESC}$ and $\theta_{DESC}$.
On the grid points, equibrium quantities required to calculate the RHS of Eq.\eqref{eq:ode1-fl} and Eq.\eqref{eq:ode2} given $\zeta$ and $Y$
are calculated and interpolated using cubic splines.

After the setup stage, Eqs. \eqref{eq:ode1-fl} and \eqref{eq:ode2} are solved using the RK45 integrator given initial condition for $\eta$ and $Y$.
To find a saturated state, two methods can be used. 
One option is a shooting method that seeks flux tubes that satisfy $\eta(\zeta_0) = \eta(\zeta_1)=0$ at the boundary of the simulation box. 
Brent's method\cite{brent_algorithms_1972-1} is used to find the value of $Y_0$ that gives $\eta(\zeta_1)=0$, where $Y_0$ is $Y$ at $\zeta=\zeta_0$.
The other method utilizes the piecewise $C^1$ equilibrium flux tubes introduced in section \ref{sec:4}.
A series of flux tubes is calculated starting with different $Y_0$ at both ends of the simulation domain and ending at 
a preset $\zeta=\zeta_{cut}$ in the middle.
On both sides of $\zeta_{cut}$, the solutions is interpolated as a function of $Y_0$ and $\zeta$, $\eta(\zeta,Y_0)$. 
When $Y_0$ is fixed and $\zeta$ varied, $\eta(\zeta, Y_0)$ is a solution to Eq.\eqref{eq:ode1-fl} and Eq.\eqref{eq:ode2} with initial condition $Y=Y_0$.
The equilibrium flux tube is found when $\partial_\zeta \eta(\zeta_{cut}^-,Y_0)=\partial_\zeta \eta(\zeta_{cut}^+,Y_0)$ with the 
piecewise $C^1$ solutions. 
The energy of the flux tube can be calculated by the variational approach introduced in section \ref{sec:4}.
\section{Variational derivation of Eq.\eqref{eq:forcebalance}}\label{ap:energy}

We want to minimize the functional in Eq. \eqref{eq:energy}
with the constraint that the variation stays on the $\alpha$ surface
\begin{align}
    \mathbf{\delta r} \cdot \grad \alpha = 0.
\end{align}
With this constraint, we have
\begin{align}\label{eq:deltar}
    \bd {\delta r} = \delta r_1\bd{  b_{in}} + \delta r_2 \bd{e_\perp}
\end{align}
where $\delta r_1$ and $\delta r_2$ are scalars and $\bd{b}_{in}=\bd{B}_{in}/|\bd{B}_{in}|$ is the unit vector 
parallel to the displaced flux tube. The displacement $\delta \bf r$ should be a continuous function of $l$.

The change in the energy functional due to the displacement $\bd{\delta r}$, after integrating by part, is
\begin{align}\label{eq:variational}
    \delta \mathcal{E} =& \int_{l_0}^{l_1} \bd{\delta r} \cd (\grad B_{in} - B_{in} \bd b_{in}\cd \grad \bd b_{in} - \bd b_{in} (\bd b_{in}\cd \grad B_{in} )) dl \nonumber\\
    +& \vb{(\bd B_{in} \cd \bd{\delta r} )}_0^1.
\end{align}

For a smooth perturbed flux tube that goes to zero at $-\infty$ and $+\infty$, Eqs. \eqref{eq:deltar} and \eqref{eq:variational} gives
\begin{align}\label{eq:forceB_parallel}
    \bd b_{in} \cd (\grad B_{in} - B_{in} \bd b_{in}\cd \grad \bd b_{in} - \bd b_{in} (\bd b_{in}\cd \grad B_{in} ) )= 0,
\end{align}
and
\begin{align}\label{eq:forceB_perp}
    \bd{e_\perp} \cd (\grad B_{in} - B_{in} \bd b_{in}\cd \grad \bd b_{in} - \bd b_{in} (\bd b_{in}\cd \grad B_{in} ) ) = 0.
\end{align}

Eq. \eqref{eq:forceB_parallel} is automatically satisfied. Eq. \eqref{eq:forceB_perp} gives
Eq. \eqref{eq:forcebalance} after some manipulations. Indeed, Using Eq. \eqref{eq:Bin}, we find
\begin{align}
    &B_{in} \bb{ \grad B_{in} - B_{in} \bd b_{in}\cd \grad \bd b_{in} - \bd b_{in} (\bd b_{in}\cd \grad B_{in} ) }\nonumber \\ 
    &= \grad (B_0^2/2 + \mu_0 p)  - \bd B_{in} \cd \grad \bd B_{in} \nonumber \\ 
    &= \bd B_0 \cd \grad \bd B_0 - \bd B_{in} \cd \grad \bd B_{in} + (\mu_0\grad p - \bd J \x \bd B).
\end{align}
Therefore, for a numerical equilibrium where $\mu_0\grad p - \bd J \x \bd B \neq 0$, there will be a mismatch between the solution of the flux tube equilibrium solution and the stationary point of the energy functional.

For section \ref{sec:4}, it is useful to consider a perturbed flux tube that goes to zero at $-\infty$ and $+\infty$ but has a discontinuity of slope at a location $l_{cut}$.
The change of energy due to a displacement $\bd{\delta r}$ of such a flux tube is 
\begin{align}\label{eq:ap:discounty}
    \delta \mathcal{E} =& \int_{-\infty}^{l_{cut}} \bd{\delta r} \cd (\grad B_{in} - B_{in} \bd b_{in}\cd \grad \bd b_{in} - \bd b_{in} (\bd b_{in}\cd \grad B_{in} )) dl \nonumber\\
    +& \int_{l_{cut}}^{+\infty} \bd{\delta r} \cd (\grad B_{in} - B_{in} \bd b_{in}\cd \grad \bd b_{in} - \bd b_{in} (\bd b_{in}\cd \grad B_{in} )) dl \nonumber \\ 
    -& \bb{\mathbf{B}_{in} }_{l_{cut}}\cd \vb{\bd {\delta r}}_{l_{cut}}  
    = -\bb{\mathbf{B}_{in} }_{l_{cut}}\cd \vb{\bd {\delta r}}_{l_{cut}}.
\end{align}
The last equality holds when the flux tube pieves on both sides of the discontinuity are in equilibrium.




\section{Behavior of the displaced flux tube at large $\zeta$}\label{ap:decay}
In this appendix, we calculate the behavior of the nonlinear ballooning equation at large $\zeta$ by performing 
a two-scale expansion of the full nonlinear ballooning equation \eqref{eq:forcebalance}.
We need a two-scale expansion because the coefficient in Eq. \eqref{eq:forcebalance} have two different
dependence on $\zeta$: some of them oscillate and some of them change secularly with $\zeta$. The 
particular term that increase secularly is $|\bd{e}_\perp|$, which can be 
conveniently expressed as
\begin{align}
    |\bd{e_\perp}|^2 = |\grad \psi |^2 L_a^2 + \frac{B_0^2}{|\grad \psi|^2}.
\end{align}
Here $L_a = \iota'(\eta) \zeta - h(\eta,\zeta)$ and 
\begin{align}
    &h = \frac{\bbr{\grad \theta - \iota \grad \zeta}\cd \grad \psi}{|\grad \psi|^2}.
\end{align}
When $\zeta$ is large ($\eta \to 0$), $L_a$ can be approximated by $L$, which is defined as
\begin{align}
    &L = \vb{\iota'}_0 \zeta - \vb{h}_0(\zeta).
\end{align}
The subscript $|_0$ indicates that the function is evaluated on an approximation to the $\alpha$ surface,
with the coordinates given by $\rho_0$, $\zeta$ and $\theta = f(\rho_0) + \iota(\rho_0) \zeta + \iota'(\rho_0) \eta(\zeta)\zeta$.


Since $|\bd{e}_\perp|$ can be approximated by $L|\grad \psi|$ at large $L$, we find it convenient to use the two 
scale expansion for $\eta$
\begin{align}
    \eta = L^{-\gamma}\bbr{ \eta_0(\zeta) + \eta_1(\zeta) L^{-1} + \eta_2(\zeta) L^{-2} \cdots }.
\end{align}
Here the functions $\eta_{i}$ with $i\ge 0$, represent the fast variation of $\eta$ with $\zeta$, and 
$L$ is the slow scale along the field line.
We assume $L\gg 1$. Note that for $\gamma \leq 1$, the displaced flux tube does not converge to the original flux tube because 
$\theta \sim \iota' \eta \zeta$ for large $\zeta$. Thus from here on we assume $\gamma > 1$.

For $\gamma > 1$ and $L \gg 1$, $\eta \sim L^{-\gamma}$ and $d\eta/d\zeta \sim  d \eta_0 /d \zeta L^{-\gamma} + \eta_0 L^{-\gamma-1}$.
Therefore, the derivative of $\eta$ with respect to $\zeta$ tends to zero even when multiplied by the large coordinate $L$, $Ld\eta/d\zeta \sim O(L^{1-\gamma})\ll 1$. This means that $|\mathbf{e_\perp}B_\perp|$ tends to zero as $L\to \infty$ and 
$B_\perp \simeq (\mathbf{B_0}\cd \grad \zeta/B_0) d{\eta}/d{\zeta}$ to leading order in $L$.
Using this fact, we can estimate the leading order behavior of different terms in the nonlinear
ballooning equation. The field line bending term $\mathbf{B}_{in}\cd \grad [(|\mathbf{e}_\perp |^2/B_{0})B_\perp]$ is $O(L^2 d\eta /d\zeta)\sim O(L^{2-\gamma})$, the instability drive term $(B_\parallel^2-1) B_{0} \boldsymbol{\kappa}\cd \mathbf{e}_\perp$ is $O(\max( L^{1-\gamma}\eta_0, L^3 (d\eta /d\zeta))^2)$
and the nonlinear local shear term $B_\perp^2 B_{0} \mathbf{e}_\perp  \cd \grad (|\mathbf{e}_\perp |^2/2B_{0}^2)$ is $O(L^3 (d\eta /d\zeta)^2)$. 
The leading order term among all the three terms is the linear field line bending term, which consists of only fast scale derivatives of $\eta_0$
\begin{align}
    \dd{}{\zeta} \frac{|\grad \psi|^2}{B_0^2} \dd{\eta_0}{\zeta}  = 0
\end{align}
This equation has a solution of the form of $d \eta_0 / d \zeta = B_0^2 C/ |\grad \psi|^2$,
where $C$ is a constant of integration. For $\eta_0$ not to change secularly with $\zeta$,
$B_0^2 C/ |\grad \psi|^2$ needs to average to zero after several toroidal turns. This is only 
possible if $C=0$, and hence we have $d \eta_0 / d \zeta = 0$.

The result $d\eta_0/d\zeta = 0$ means that we have $d\eta/d\zeta \sim  O(L^{-\gamma-1}) $. The nonlinear local shear term $B_\perp^2 B_{0} \mathbf{e}_\perp  \cd \grad (|\mathbf{e}_\perp |^2/2B_{0}^2)$ 
is $O(L^{1-2\gamma}) \ll O(L^{-\gamma})$.
One can also show that the leading order nonlinear contribution in both the field line bending term $\mathbf{B}_{in}\cd \grad [(|\mathbf{e}_\perp |^2/B_{0})B_\perp]$ and the instability drive term $(B_\parallel^2-1) B_{0} \boldsymbol{\kappa}\cd \mathbf{e}_\perp$
are $O(L^{1-2\gamma})$ or smaller. Therefore, the contributions $O(L^{1-\gamma})$ and $ O(L^{-\gamma})$
are all from linear terms. This leads to the same results that one obtains expanding the linear ballooning equation at large $\zeta$ (see \cite{connor_ballooning_1987}) and hence predicts same $\gamma$. This is the $\gamma$ provided by Mercier analysis.

To summarize, even for the fully nonlinear ballooning equation, the behavior of the displaced flux tube is predicted by 
the exponent $\gamma$ from the Mercier analysis of the linear ballooning equation when $\gamma > 1$.
As we explained above, the solutions for $\gamma \ge 1$ are not physical.


\begin{thebibliography}{10}
\expandafter\ifx\csname url\endcsname\relax
  \def\url#1{{\tt #1}}\fi
\expandafter\ifx\csname urlprefix\endcsname\relax\def\urlprefix{URL }\fi
\providecommand{\eprint}[2][]{\url{#2}}

\bibitem{ham_nonlinear_2018}
Ham C~J, Cowley S~C, Brochard G {\em et~al.\/} 2018 {\em Plasma Physics and
  Controlled Fusion\/} {\bf 60} 075017 ISSN 0741-3335 publisher: IOP Publishing

\bibitem{ham_nonlinear_2016}
Ham C, Cowley S, Brochard G {\em et~al.\/} 2016 {\em Physical Review Letters\/}
  {\bf 116} 235001 publisher: American Physical Society

\bibitem{hirsch_major_2008}
Hirsch M, Baldzuhn J, Beidler C {\em et~al.\/} 2008 {\em Plasma Physics and
  Controlled Fusion\/} {\bf 50} 053001 ISSN 0741-3335, 1361-6587

\bibitem{sakakibara_mhd_2008}
Sakakibara S, Watanabe K~Y, Suzuki Y {\em et~al.\/} 2008 {\em Plasma Physics
  and Controlled Fusion\/} {\bf 50} 124014 ISSN 0741-3335, 1361-6587

\bibitem{klinger_overview_2019}
Klinger T, Andreeva T, Bozhenkov S {\em et~al.\/} 2019 {\em Nuclear Fusion\/}
  {\bf 59} 112004 ISSN 0029-5515, 1741-4326

\bibitem{weller_significance_2006}
Weller A, Sakakibara S, Watanabe K~Y {\em et~al.\/} 2006 {\em Fusion Science
  and Technology\/} {\bf 50} 158--170 ISSN 1536-1055, 1943-7641

\bibitem{leonard_edge-localized-modes_2014}
Leonard A~W 2014 {\em Physics of Plasmas\/} {\bf 21} 090501 ISSN 1070-664X

\bibitem{connor_high_1979}
Connor J~W, Hastie R~J and Taylor J~B 1979 {\em Proceedings of the Royal
  Society of London. A. Mathematical and Physical Sciences\/} {\bf 365} 1--17
  publisher: Royal Society

\bibitem{geiger_equilibrium_2004}
Geiger J~E, Weller A, Zarnstorff M~C {\em et~al.\/} 2004 {\em Fusion Science
  and Technology\/} {\bf 46} 13--23 ISSN 1536-1055, 1943-7641

\bibitem{ohdachi_observation_2017}
Ohdachi S, Watanabe K, Tanaka K {\em et~al.\/} 2017 {\em Nuclear Fusion\/} {\bf
  57} 066042 ISSN 0029-5515, 1741-4326

\bibitem{connor_shear_1978}
Connor J~W, Hastie R~J and Taylor J~B 1978 {\em Physical Review Letters\/} {\bf
  40} 396--399 ISSN 0031-9007

\bibitem{greene_second_1981}
Greene J~M and Chance M~S 1981 {\em Nuclear Fusion\/} {\bf 21} 453 ISSN
  0029-5515

\bibitem{dewar_ballooning_1983}
Dewar R~L and Glasser A~H 1983 {\em Physics of Fluids\/} {\bf 26} 3038

\bibitem{hegna_stability_1998}
Hegna C~C and Nakajima N 1998 {\em Physics of Plasmas\/} {\bf 5} 1336--1344
  ISSN 1070-664X

\bibitem{sanchez_ballooning_2000}
Sanchez R, Hirshman S~P, Ware A~S {\em et~al.\/} 2000 {\em Plasma Physics and
  Controlled Fusion\/} {\bf 42} 641--653 ISSN 0741-3335, 1361-6587

\bibitem{miller_hot_1987}
Miller R~L and Dam J~W~V 1987 {\em Nuclear Fusion\/} {\bf 27} 2101 ISSN
  0029-5515

\bibitem{anderson_methods_1990}
Anderson D, Cooper W, Gruber R {\em et~al.\/} 1990 {\em The International
  Journal of Supercomputing Applications\/} {\bf 4} 34--47 ISSN 0890-2720

\bibitem{cowley_explosive_1997}
Cowley S~C and Artun M 1997 {\em Physics Reports\/} {\bf 283} 185--211 ISSN
  0370-1573

\bibitem{hurricane_nonlinear_1997}
Hurricane O~A, Fong B~H and Cowley S~C 1997 {\em Physics of Plasmas\/} {\bf 4}
  3565--3580 ISSN 1070-664X

\bibitem{henneberg_interacting_2015}
Henneberg S~A, Cowley S~C and Wilson H~R 2015 {\em Plasma Physics and
  Controlled Fusion\/} {\bf 57} 125010 ISSN 0741-3335 publisher: IOP Publishing

\bibitem{zhu_ballooning_2008}
Zhu P and Hegna C~C 2008 {\em Physics of Plasmas\/} {\bf 15} 092306 ISSN
  1070-664X

\bibitem{zhou_approach_2021}
Zhou Y, Ferraro N, Jardin S {\em et~al.\/} 2021 {\em Nuclear Fusion\/} {\bf 61}
  086015 ISSN 0029-5515, 1741-4326

\bibitem{zhou_benign_2024}
Zhou Y, Aleynikova K, Liu C {\em et~al.\/} 2024 {\em Physical Review Letters\/}
  {\bf 133} 135102 ISSN 0031-9007, 1079-7114

\bibitem{panici_desc_2023}
Panici D, Conlin R, Dudt D~W {\em et~al.\/} 2023 {\em Journal of Plasma
  Physics\/} {\bf 89} 955890303 ISSN 0022-3778, 1469-7807

\bibitem{nelson_design_2002}
Nelson B, Berry L, Brooks A {\em et~al.\/} 2002 Design of the {National}
  {Compact} {Stellarator} {Experiment} ({NCSX}) core {\em Proceedings of the
  19th {IEEE}/{IPSS} {Symposium} on {Fusion} {Engineering}. 19th {SOFE} ({Cat}.
  {No}.{02CH37231})\/} (Atlantic City, NJ, USA: IEEE) pp 285--289 ISBN
  978-0-7803-7073-9

\bibitem{brent_algorithms_1972-1}
 1972 Algorithms for {Minimization} without {Derivatives} {\em Algorithms for
  minimization without derivatives\/} Prentice-{Hall} series in automatic
  computation (Englewood Cliffs, N.J: Prentice-Hall) pp Ch. 3--4 ISBN
  978-0-13-022335-7

\bibitem{cowley_explosive_2015}
Cowley S~C, Cowley B, Henneberg S~A {\em et~al.\/} 2015 {\em Proceedings of the
  Royal Society A: Mathematical, Physical and Engineering Sciences\/} {\bf 471}
  20140913 publisher: Royal Society

\bibitem{newcomb_hydromagnetic_1960}
Newcomb W~A 1960 {\em Annals of Physics\/} {\bf 10} 232--267 ISSN 00034916

\bibitem{drevlak_estell_2013}
Drevlak M, Brochard F, Helander P {\em et~al.\/} 2013 {\em Contributions to
  Plasma Physics\/} {\bf 53} 459--468 ISSN 0863-1042, 1521-3986

\bibitem{connor_ballooning_1987}
Connor J~W and Taylor J~B 1987 {\em The Physics of Fluids\/} {\bf 30}
  3180--3185 ISSN 0031-9171 publisher: AIP Publishing

\end{thebibliography}
\end{document}